\documentclass[manuscript,table,screen,acmsmall]{acmart}

\usepackage{booktabs} 
\usepackage{graphicx}
\usepackage{multirow}
\usepackage{hhline}
\usepackage{float}
\usepackage{subcaption}
\usepackage{algorithm}
\usepackage[noend]{algpseudocode}

\newenvironment{myquote}[1]%
  {\list{}{\leftmargin=#1\rightmargin=#1}\item[]}%
  {\endlist}

\AtBeginDocument{%
  \providecommand\BibTeX{{%
    \normalfont B\kern-0.5em{\scshape i\kern-0.25em b}\kern-0.8em\TeX}}}

\setcopyright{acmcopyright}
\acmJournal{TOIS}
\acmYear{2020} \acmVolume{4} \acmNumber{TOIS} \acmArticle{1} \acmMonth{4}

\begin{document}

\title{Hierarchical Knowledge Graphs: A Novel Information Representation for Exploratory Search Tasks}


\author{Bahareh Sarrafzadeh}
\affiliation{%
 \institution{Cheriton School of Computer Science,
			University of Waterloo}
  \city{Waterloo}
  \country{Canada}
}
\email{bsarrafz@uwaterloo.ca}

\author{Adam Roegiest}
\affiliation{%
 \institution{Kira Systems}
  \city{Toronto}
  \country{Canada}
}
\email{adam.roegiest@kirasystems.com}

\author{Edward Lank}
\affiliation{
\institution{Cheriton School of Computer Science,
			University of Waterloo}
  \city{Waterloo}
  \country{Canada}
}
\email{lank@uwaterloo.ca}

\renewcommand{\shortauthors}{Sarrafzadeh, Roegiest, and Lank}

\begin{abstract}
  In exploratory search tasks, alongside information retrieval, information representation is an important factor in sensemaking. 
  In this paper, we explore a multi-layer extension to knowledge graphs, hierarchical knowledge graphs (HKGs), that combines hierarchical and network visualizations into a unified data representation as a tool to support exploratory search. 
We describe our algorithm to construct these visualizations, analyze interaction logs to quantitatively demonstrate performance parity with networks and performance advantages over hierarchies, and synthesize data from interaction logs, interviews, and thinkalouds on a testbed data set to demonstrate the utility of the unified hierarchy+network structure in our HKGs.
Alongside the above study, we perform an additional mixed methods analysis of the effect of precision and recall on the performance of hierarchical knowledge graphs for two different exploratory search tasks. While the quantitative data shows a limited effect of precision and recall on user performance and user effort, qualitative data combined with post-hoc statistical analysis provides evidence that the type of exploratory search task (e.g., learning versus investigating) can be impacted by precision and recall. Furthermore, our qualitative analyses find that users are unable to perceive differences in the quality of extracted information. We discuss the implications of our results and analyze other factors that more significantly impact exploratory search performance in our experimental tasks\footnote{This paper is an extension of our prior work \cite{sarrafzadeh17}.}.
\end{abstract}



\maketitle

\section{Introduction}
Information Retrieval (IR) research explores a wide range of questions connected to the goal of helping a user find useful information in response to a need \cite{Allan05}. Given this goal and the availability of on-line information, it is unsurprising that significant IR research effort exists in supporting web search \cite{broder2002,  marchionini06, wilson10, white09}.  It is well-established that, within the domain of web search, there exists two broad categories of search: look-up searches, which leverage a ``search by query'' strategy for information seeking
; and browsing (i.e., ``search by navigation'') \cite{olston03,jul97, marchionini2019}. Our interest is primarily in the second category of search, and, in particular, in the broad category of information seeking that has been classified as \emph{Exploratory Search} where the goal involves ``learning" (i.e., developing new knowledge) or ``investigating" (i.e., applying analysis, synthesis, and evaluation), rather than simply ``looking-up"~\cite{marchionini06}.

The 2018 SWIRL Workshop \cite{swirl2018} emphasizes the need for research in supporting complex, evolving and exploratory information seeking goals.  Supporting these goals requires advances in the algorithms that provide information, the interfaces that represent this information, and the evaluation methods that support these goals. 
To these ends, this paper focuses on two aspects that are critical for success in exploratory search tasks:
challenges in the algorithms that are used to identify and extract information that is relevant to a searcher's query;
and, on the interfaces that visualize the output of these algorithms and facilitate a searcher's interaction with and exploration of the retrieved information.

Prior work indicates that interfaces supporting exploratory search benefit from both ``search by query'' and ``search by navigation'' paradigms~\cite{olston03,jul97, marchionini2019},
and that such interfaces require concise and understandable representations of search results that accommodate a searcher's `state of knowledge' and allow the searcher to identify promising directions for exploration (i.e., `information scent' \cite{pirolli1997}) to guide the user during navigation \cite{hearst09, pirolli2000, olston03, sarrafzadeh16, sarrafzadeh17}.
This desire to support navigation has given rise to interfaces that represent the structures of information, which, broadly speaking, represent information as hierarchies or as networks~\cite{chen98}. In hierarchical representations, information is represented using categorical labels, such as those present in faceted browsing and automatic clustering  of research results. On the the other hand, network representations typically manifest as entity relationships in a graph-like structure. Rather than clustering similar objects, the connections between objects (e.g., people, places, and things) in a document collection are used to represent a relationship between two (or more) entities. Such network representations have been seen in the use of knowledge graphs~\cite{james91} and concept maps ~\cite{novak08}.

In recent work, we~\cite{sarrafzadeh16} explored the relative benefits of hierarchies and networks and noted that the benefits are largely complementary: hierarchies provide users with some understanding of central topics, allowing them to develop a better overview of information; whereas networks allow people to glean concrete information from the representation rather than needing to extensively read individual documents. 
Given the complementary advantages of
knowledge graphs and hierarchies, the first research question in
this paper explores \emph{whether we can algorithmically generate a seamless data structure that combines the advantages of both hierarchies
and networks.  Is it possible to combine the benefits of data overviews of hierarchies and the low-level information content of networks in a unified representation of search results?}

To explore this question, we design a new structured representation of information, hierarchical knowledge graphs (HKGs - Section \ref{HKG}), that are extracted from a document collection which is generated by a search query, and then evaluate the efficacy of HKGs as a combined representation of low-level entity relationships and high-level central concepts (Section \ref{HKG_Evaluation_PartI}). To generate HKGs, we use entity-relationship tuples generated by a simple parsing
algorithm~\cite{sarrafzadeh13_techReport} and then manually corrected to ensure accurate information extraction.  We create hierarchies within these knowledge graphs using a dynamic thresholding approach described in this paper (Section \ref{HKG_Creation}). Finally, we evaluate these HKGs using a mixed methods
approach. Quantitative data argues that HKGs preserve the transparency advantages of knowledge graphs and structural advantages
of hierarchies (Section \ref{Results_Quantitative_PartI}). Qualitative data triangulates with quantitative observations and provides additional insight into the advantages and
disadvantages of both hierarchical and network visualizations (Section \ref{Results_Qualitative_PartI}).  



While our qualitative and quantitative data indicate that HKGs provide benefits of both knowledge graphs and hierarchies, we note that, to determine this benefit, we assumed that information extraction was performed accurately. However, information extraction (IE) algorithms are not perfect \cite{stanovsky16}, meaning that the base level information to be presented to a user may contain errors and omissions, and, as such, may reflect compromises in either \emph{precision} or \emph{recall} by the algorithm. The TAC~\footnote{https://tac.nist.gov/} and TREC~\footnote{https://trec.nist.gov/} tracks on evaluating IE and question answering (QA) algorithms report varying levels of precision and recall for these algorithms when compared against manually created ground truth datasets. 
Accordingly, given that IE algorithms are not perfect, an additional research question is what happens if there are errors in the extraction, i.e., \emph{how resilient are representations of search results such as HKGs to these errors? And how do errors impact a user's ability to leverage these representations to acquire knowledge?}

To explore this second research question, we present a second mixed-methods study \cite{creswell2013research} that examines the impact of imperfect IE on exploratory search via HKGs (Section \ref{HKG_Evaluation_PartII}).  Users interact with two HKGs, one manually corrected by human experts (the typical approach to generating ground truth to benchmark IE systems \cite{stanovsky16}) and a second graph that is automatically generated and exhibits significantly lower precision and recall compared to the manually generated graph. 

While our expectation was that precision and recall would impact user performance (i.e., success in an exploratory search task) or the effort expended during search (e.g., the number of documents viewed), our results indicate that neither performance nor effort was significantly impacted by differing levels of precision and recall (Section \ref{Results_Quantitative_PartII}). To probe this result in greater detail, we analyze qualitative data collected via observations and interviews.  Our qualitative data indicates that task characteristics may be an important factor to consider in exploratory search (Section \ref{Results_Qualitative_PartII}).  Specifically, for investigate-style tasks \cite{marchionini06} where there is a defined set of facts to retrieve, recall may impact user behavior because it is necessary to find specific facts within the information presented.  In contrast, more open-ended learn/comprehension or comparison tasks, where salient data can be more flexibly applied by the user, seem more resilient to lower recall rates.  

Overall, this paper makes two contributions to IE.  First, it demonstrates the benefits of hierarchies and networks as complementary representations of information in support of exploratory search.  It does this through the design and evaluation of a novel data structure, hierarchical knowledge graphs, which combine both low level entity relationships and hierarchical depictions of more centrally connected entities within a document corpus.  Second, it presents an in situ evaluation of hierarchical knowledge graphs given a realistic information extraction algorithm. If one accepts that the goal of improved IE is to substantively improve task outcomes, then the fact that we have found only one instance of IE evaluation that considers user performance \cite{chu07} is a significant concern, and this paper highlights how different exploratory search task types (e.g., investigate vs compare/comprehension) may be more or less resilient to varied precision and recall.

\section{Background}
\label{Background}
While there has been research on understanding complex and exploratory search (see \cite{white09, wilson10} for a survey), there are many open questions when it comes to the design and evaluation of IR systems that provide tailored and adaptive support for different search tasks.
Given our interest in exploratory search, in this section we first survey three areas of past research that explore support for users with more complex, exploratory search tasks.
First, there has been a growing body of work in the IR community that aims to deliver ``information'' and not documents. Within this body of work, Open Information Extraction (Open IE) \cite{banko07} techniques have been widely applied to extract semantic information from the text of documents.
The second area focuses on investigating ways that search systems can represent and provide the extracted information to help searchers in evaluating and contextualizing search results. 
Finally, developing solutions to support users' exploratory search tasks also includes significant challenges in evaluation. 


Alongside research on supporting exploratory search, our second research question probes the effect of error on information seeking tasks. It is often assumed if an evaluation measure coupled with a test collection reveals that system A provides higher quality output than system B, then the user will both prefer system A and that system A will more effectively support the user's information seeking task \cite{Allan05, vakkari12}.  However, in our analysis of IR research on this topic, we have found that the relationship between output quality and system efficacy is not clear.  This ambiguity is highlighted in the second part of this Background section.

\subsection{System Support for Exploratory Search}
\label{SystemSupport}
To support exploratory search, a system must have two components: an information extraction system that can identify and extract relevant information from a corpus (e.g., search engine results); and, an interactive UI that presents the information to users and allows users to browse the extracted information for sensemaking~\cite{jul97, olston03, marchionini2019} (e.g. a browsable list of documents retrieved).
Any arbitrary system designed to support exploratory search will exist on a range from a standard search engine interface (an ordered list where a user can select and browse individual documents) to systems that extract and synthesize information for the user. An example of the latter type of system includes those found at the TREC Complex Answer Retrieval track~\cite{dietz17trec}, which explores the design of systems that apply information extraction to synthesize content and generate an essay on a particular topic. Ultimately, any system on this spectrum requires the two components identified above: an information extraction component and a UI to support the user in acquiring the desired information.

A common approach to present-day exploratory search systems is to leverage IE to identify relevant entities and their relationships within the retrieved documents. These entities and relationships can then be displayed in graphical form (e.g., as a knowledge graph, a hierarchy, a concept map) which provides the user with a spatial representation of the information space for sensemaking~\cite{pirolli2000}. This representation can then be embedded in an interface that allows a user to interact with the representation, to filter and select specific content, and essentially to explore the information returned~\cite{sarrafzadeh16}. 
As a main focus of this paper is on designing an interface that provides an effective representation of search results for exploratory search tasks, we review the existing organizations of search results next.

\subsubsection{Organizing Search Results}
Information seekers often express a need for tools that organize search results into meaningful constructs in order to support sensemaking and navigation \cite{hearst06, marchionini06}.
Because of the importance of structure in search, there have been efforts to contrast strengths and weaknesses of different spatial representations and groupings of search results. A taxonomy of techniques for organizing
search results was proposed by Wilson et al. \cite{wilson10}. They identify
two main classes of approaches: 
(1) adding classifications, i.e. coupling results with additional metadata and classifications such that searchers can interact and control the presentation
of results (e.g., faceted browsing or categories); and (2) providing alternative
or complementary representations of search results (e.g., a network representation).  Ltifi et al. \cite{ltifi2009}, as with Wilson et al., present three alternative representations of search results: vectorial lists, hierarchies, and networks. Modern day search engines produce an ordered list of documents retrieved, an example of a vectorial list; hierarchies and networks echo Wilson's taxonomy of search results representations (i.e., adding hierarchical classifications and complementary representations of information, respectively).

Looking first at structured classification, early forays into the domain of structuring search results  contrasted categories with automatic clustering to support search. Hearst \cite{hearst99} showed that categories, because they were more interpretable for the user, captured important information about the document but became unwieldy when the document corpus was too large. Clusters, by comparison, were highly variable with respect to quality and were often less meaningful for the user.

Given the lack of intuitiveness associated with clustering \cite{hearst06} and a desire for understandable hierarchies in which categories are presented at uniform levels of granularity \cite{pratt99, rodden01}, alongside specified hierarchies such as tables-of-contents, researchers have explored faceted categories, i.e. categories that are semantically related to the search task of the user, to organize search results. These include systems that define faceted categories \cite{yee03}, research that studies the use of facets to support browsing \cite{capra07}, and research that identifies strengths and weaknesses of faceted browsers \cite{wilson09}. In terms of strengths and weaknesses, faceted browsing has proven beneficial for users already clear about their search task \cite{wilson09}; additional information on interactions between facets (e.g. inter-facet relationships) is helpful when users are unfamiliar with a domain and need `sensemaking'. In other words, exploratory tasks (e.g. learning or investigating \cite{marchionini06}) are precisely those tasks where interactions between facets are needed.

The need to represent relationships between facets or concepts has given rise to the use of network structures to depict relationships between concepts or entities in a corpus. These network structures include concept maps, knowledge graphs, and other entity-relationship diagrams. Concept mapping has been widely used in education as a method for knowledge examination, sharing, and browsing \cite{novak08, carnot03}. Knowledge graphs have been popularized by Google to represent web-based information. One drawback to network structures is it is hard both to get an overview of an information network and to navigate through the network effectively: users are easily ``lost'' in these systems \cite{cockburn96,komlodi07,olston03}.

A final question within this space is how competing representations fare in presenting results for exploratory search tasks. While some past research has explored using questionnaires to determine the efficacy of different knowledge representations \cite{novick01}, or has evaluated the efficacy of hierarchies, networks, or concept maps with respect to ordered lists (e.g. \cite{capra07, norman88, ducheneaut01,
sarrafzadeh14, amadieu10, amadieu14, 
carnot03}), we have found little research that directly compares representations such as networks and hierarchies to understand their competing affordances. The one exception to this is our recent work \cite{sarrafzadeh16} on contrasting the efficacy of network and hierarchical representations of search results in supporting information seeking tasks. We find that networks eliminate the need for reading documents -- users can glean information from the networks with statistically significantly less time spent reading -- and that hierarchies particularly benefit low-knowledge participants by giving them an effective overview of the domain.

Given our past observation of the complementary benefits of hierarchies and networks, one question is whether studies have examined the use of hierarchies and networks as combined -- synchronized and simultaneous -- representations of search results. To the best of our knowledge, this paper represents the first such attempt; part of the challenge may arise from the complexity of seamlessly integrating both hierarchies and networks into a single unified structure. For example, hierarchies are typically best when the structure aligns well with the user's task, but, given this alignment, entities in networks may have many multiple `parents' within the structure, yielding a many-to-many relationship within the hierarchy, i.e. a three-dimensional graph.  

In section \ref{HKG}, we propose a solution to this challenge by synthesizing, algorithmically, a hierarchical structure from the a low-level network representation.  We then evaluate this novel representation of search results in Section \ref{HKG_Evaluation_PartI}.  This, naturally, leads to the need to understand related work on how best to evaluate systems to support exploratory search.

\subsection{Evaluating Exploratory Search Systems}
\label{EvaluatingESS}
Given a system to support exploratory search, we must determine how best to characterize the performance of an exploratory search system.
There are two aspects to system performance: accuracy and effectiveness. 
Assessing the accuracy of an algorithm can be performed through benchmarking and/or combined efforts tasks (e.g., TREC or CLEF tasks). System effectiveness for exploratory search, on the other hand, requires evaluating how well the systems aids in the exploratory search tasks it is designed around. In the following two subsections, we describe, in more detail, these two types of assessments. 

\subsubsection{System Accuracy}
Evaluating information extraction is challenging.  There are no clear guidelines as to what constitutes a valid proposition to be extracted, and most information extraction evaluations  consist of a post-hoc manual evaluation of a small output sample~\cite{stanovsky16}.  There is also no agreement on an appropriate data set to use for information extraction
~\cite{niklaus2018survey}.
However, Stanovsky and Dagan \cite{stanovsky16} have developed a methodology that leveraged the recent formulation of QA-SRL~\cite{he05} to create the first independent and large-scale gold benchmark corpus.\footnote{The corpus is available at: https://github.com/gabrielStanovsky/oie-benchmark} Stanovsky and Dagan's benchmark is based on a set of guiding principles that underlie most Open IE approaches. This benchmark has provided an opportunity to evaluate the output of an Open IE system using both precision and recall.  While acknowledging that this benchmark may not be perfect, in this paper we leverage it as the most up-to-date standard for evaluating IE systems.

\subsubsection{System Effectiveness}
It has long been understood in IR that a system understanding of relevance is not always consistent with what a user desires~\cite{saracevic} and so we must also understand how systems impact user performance.  
The recent SWIRL Workshop \cite{swirl2018} has identified the most relevant research questions to be addressed in order to develop new evaluation models that are suited for complex and exploratory information seeking. A major step towards this goal is to design and study characteristics of search tasks that elicit exploratory behavior.  
These studies, in turn, provide data on searchers performing these tasks, specifically focused on task outcomes and searcher behaviors. 
Designing tasks for exploratory search studies can be especially difficult since inducing exploratory style search requires the searcher to individually interpret the tasks, results, and their relevance ~\cite{kules08shneiderman} which is at odds with maintaining some level of experimental control and consistency~\cite{kules08}.

To aid in the creation of appropriate exploratory search tasks, we look to Marchionini~\cite{marchionini06}, referencing Bloom's taxonomy of educational objectives~\cite{bloom56}, who distinguishes three broad categories of search tasks as {\it Lookup}, {\it Learn}, and {\it Investigate}. While these categories are depicted as overlapping activities, exploratory search is more pertinent to the {\it Learn} and {\it Investigate} activities.  As a result,  exploratory search is defined as searching that supports learning, investigating, comparing or discovering~\cite{kules08, white09}. 
From this understanding, we can distill exploratory search tasks into fitting into one of two themes. 
The first theme includes those tasks that facilitate learning to achieve knowledge acquisition, comprehension of concepts, interpretation of ideas and comparison or aggregation of concepts. The second theme covers those investigative tasks that involve discovery, analysis, synthesis and evaluation. 

Based upon the aforementioned works and a survey of existing classifications by Li and Belkin~\cite{li_belkin08}, we believe that exploratory search tasks should: provide uncertainty and ambiguity about the information need and in how to satisfy it; suggest a specific knowledge acquisition, comparison or discovery task; be in an unfamiliar domain for the searcher; represent a situation that a user can relate to and identify with; be of sufficient interest to test users; and, be formulated such that the user has enough imaginative context to facilitate immersion in the task. Any task that meets these criteria provides sufficient complexity that the end-to-end experience with an exploratory search system can be fully and properly assessed.

\subsubsection{Impact of Accuracy on Effectiveness}

As we note in the beginning of this section, there is an assumption that system accuracy (i.e., the quality of results returned) should correlate with system effectiveness (i.e., how well the system supports the task for which it was designed).  However, the correlation between accuracy and effectiveness is somewhat ambiguous in information seeking.

In document retrieval, there is a long history of research that examines how human search performance varies with system effectiveness \cite{hersh2000, turpin01, Allan05, turpin06, church90, almaskari08good, smith08, sanderson10, smucker08, wang15}. The broader goal of this line of work is to understand when system effectiveness improvements are meaningful or useful in improving users' information seeking abilities in practice. 
These studies, however, have resulted in contradictory findings. Early studies suggest that ``better'' systems, as measured by system oriented metrics (e.g., precision and recall), do not necessarily translate into better task performance \cite{hersh2000, turpin01, hersh02, turpin06, smith08}.  More recent work has detected potential correlations between various system effectiveness metrics and human preferences \cite{almaskari08good, sanderson10} and between precision and user performance \cite{smucker10}. 

Even if one gives credence to more recent work relating  effectiveness to user preference and user performance, the reason for inconsistent effects of system performance on human performance in past work is unclear.  Differences could lie in the definition of `relevance' and how it is used as a basis for evaluation of document retrieval systems \cite{doyle63, swanson77, saracevic}; they could also lie in the discrepancies between the metrics used for evaluation (e.g., MAP vs P@10) and the type of task the user performs (e.g., recall-based or complex information seeking) \cite{turpin06}; sample size may provide increased power to discriminate effects \cite{almaskari08good, sanderson10}; or, finally, differences could be a result of the lack of UI support for meaningful user interaction with the retrieved results \cite{lin08, smith08, turpin01}. Hersh et al.~\cite{hersh02} also found that while precision and recall weren't associated with success in medical QA tasks, other factors including experience of the searcher and cognitive abilities in spatial visualization were correlated with the ability to answer questions correctly.


Within the broad domain of the impact of error in IE, we were able to identify only one work by Chu-Carrol and Prager \cite{chu07} that examines how user performance degrades in the face of imperfect named entity and relation extraction.  Their results focus on assessment of document retrieval, not on assessment of support for exploratory search. Experimental results demonstrated that significant document retrieval gain can be achieved when state-of-the-art IE systems are used and that recall has more significant impact on document retrieval performance than precision when adopting the MAP metric.


Synthesizing past research, we see ambiguity in the effect of errors in the domain of information retrieval. Coupled with this, we note that exploratory search tasks require systems that support browsing, and, within information retrieval, this has given rise to systems that retrieve and present information to users in formats that support browsing.  Absent from past research is assessment of the effect of information extraction errors on exploratory search interfaces.  While we concur that IE extraction systems would ideally have perfect precision and recall, in the near term it seems unlikely that computational information extraction will be perfected, further motivating exploration of the effect of information extraction errors on interfaces that support exploratory search.
Possibly due to the ambiguous link between system performance and effectiveness, there have been calls to extend evaluation of IR systems from an analysis of the output of the system to the outcome of the search task \cite{kelly09, vakkari12}.  Furthermore, there is also an evolving drive toward evaluations of how effectively IR systems support complex, evolving, long term information seeking goals, such as learning and exploration \cite{swirl2018}.
\section{Hierarchical Knowledge Graphs}
\label{HKG}
In this section, we describe hierarchical knowledge graphs, an extension of knowledge graphs that include hierarchical information about the lower level graphical structures. Our past work \cite{sarrafzadeh16} argues that hierarchies provide a breadth-first exploration of the information allowing the user to iteratively reduce confusion, obtain an overview, and slowly exploit detail (i.e., they provide a structured way to navigate from more general concepts to more fine grained data) and are valuable when people feel a need to orient themselves; whereas network structures allow users to glean more information from the representation (document reading time is reduced), are more engaging, yield more control over exploration at the lower level of inter-concept relationships \cite{sarrafzadeh16}, and are more similar to one's mental model \cite{chen98,sarrafzadeh16, novak90,ausubel68}.  What is less clear is how best to combine hierarchies and networks in a seamless data structure, how to reliably generate hierarchies algorithmically without human intervention, and whether a data structure that combines both hierarchical and network views would yield complementary benefits.


In this section, the question we address is how to design a representation that can seamlessly merge these two representations. We take the approach that a knowledge graph will be an appropriate low-level representation and seek to incorporate a hierarchical view of this low-level representation of corpus content. To incorporate a hierarchical view into a knowledge graph, we need to find answers to the following three design questions (DQs):
\begin{enumerate}
\item How do we integrate network and hierarchical views into a single, seamless data structure?
\item How can both the global and the local view of a knowledge graph be co-visualized?
\item How can transitions between views be designed to maximize visualization stability?
\end{enumerate}
To answer these DQs, we first focus on DQ1 and describe the design of our data structure. Next, to address DQ2 and DQ3 we describe an interface that supports interaction with the data structure. Alongside our DQs, we add one additional constraint to our design. We want to ensure that both the low-level knowledge graph and the hierarchies gleaned from that knowledge graph can be automatically generated from a targeted search performed by the user.

\subsection{Visualization Design and Creation}
\label{HKG_Creation}
As noted above, given that we take the approach that a knowledge graph will constitute the lower-level visualization of our data, the task becomes creating a knowledge graph and creating a hierarchy that is gleaned from and corresponds directly to the underlying knowledge graph.

\begin{figure}[htb!]
\centering
\includegraphics[width=0.60\textwidth]{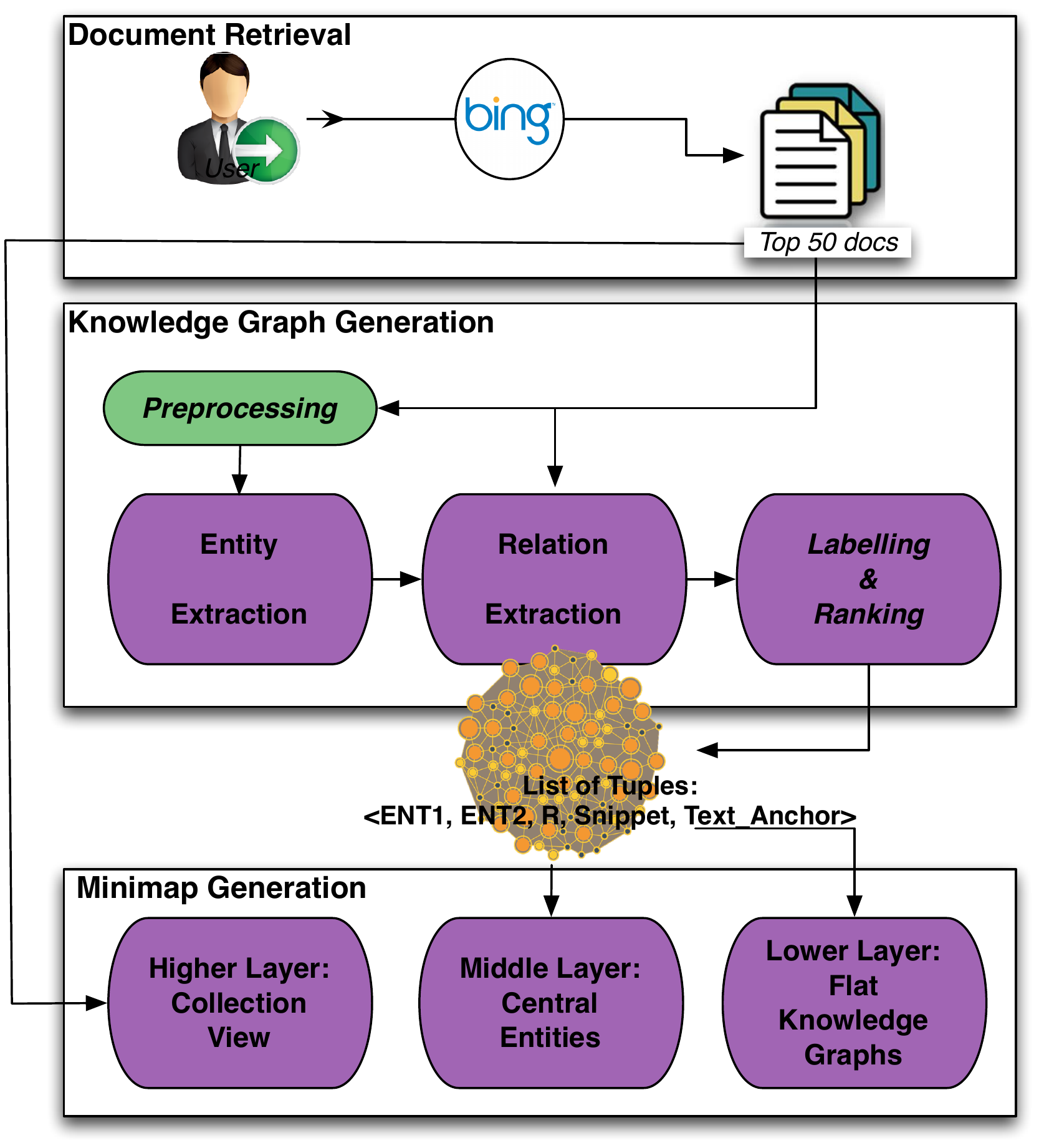}
\caption{Generating Hierarchical Knowledge Graphs}
\label{Architecture}
\vspace{-2mm}
\end{figure}

Figure \ref{Architecture} depicts the system architecture that supports the process of automatically generating the hierarchical knowledge graph representation. To simplify hierarchy generation, we create a 3-level hierarchy for any document corpus. Beyond the base layer knowledge graph, there is an intermediate layer of central concepts gleaned from the knowledge graph. Finally, at the top-level, the documents, themselves, represent the top level of the hierarchical knowledge graphs. In Figure \ref{Architecture}, three main steps are depicted to generate hierarchical knowledge graphs: Document Retrieval (yielding the top-level of the hierarchy), Knowledge Graph Generation (yielding the bottom level of the hierarchy), and Hierarchy-from-graph Generation (yielding an intermediate view of an individual knowledge graph, which we dub a \emph{minimap}\footnote{The term minimap is drawn from the gaming literature. It represents a less detailed overview of a gaming world, allowing the user to orient themselves.}).

\subsubsection{Document Retrieval.}
The Document Retrieval component aims at creating an initial document collection based on a user's query. This collection will then be used as an input for the Knowledge Graph Generation component and will represent the top view of the target hierarchy. 

To generate a document corpus, we use the Bing Search engine to retrieve the top n documents for a query while attempting to ensure a reasonable quality of information in the retrieved documents. By default, to ensure that retrieved documents are consistent in their credibility and coverage, we specify Wikipedia as the target domain. Furthermore, because it is known that searchers typically view only a few results \cite{jansen03} and rarely stray past the first
page of results \cite{beitzel04}, we selected n=10 documents to generate collections. The target domain from which to glean documents (e.g., a user might specify WebMD \footnote{https://www.webmd.com/} for medical documents, `gov' for U.S. public policy documents, `BBC' for news) and the size of the initial collection can be specified by the user at the time of query submission.
Finally, since most exploratory search tasks require multiple queries to retrieve documents for different aspects of the information need, this component assigns one partition per query so the user can narrow down the retrieved collection further.

\subsubsection{Knowledge Graph Generation.}
\label{KnowledgeGraphGeneration}
To create our knowledge graph, we designed an Open Information Extraction system that processes a text collection and generates (entity-relation-entity) triples \cite{KG4ES_tech}. This module is implemented in four phases.  During the first phase we create the input corpus by collecting retrieved documents based on a given query. Next, we extract entities from text using state-of-the-art entity taggers \footnote{https://cogcomp.cs.illinois.edu/page/software\_view/NETagger}. We then select the sentences that contain at least two entities in them and parse them using Stanford Dependency Parser. For each sentence, we extract meaningful relations between the entities by finding the shortest path in the corresponding parse tree. 
For example we extract 
``\emph{The Constitutional Act divided the Province of Quebec into Upper and Lower Canada}'' as a relationship between the entities ``\emph{Constitutional Act}'' and ``\emph{Upper Canada}.''
We constructed a set of patterns based on dependency triples that lead to semantically meaningful relations. In the final phase, we generate labels for the extracted relations and rank them based on relevance to the query and the informativeness of the extraction. 

The outcome is a set of tuples in the form of $<$entity1, entity2, relation, snippet, document\_anchor$>$. These tuples collectively correspond to a knowledge graph representation of retrieved documents where \textit{entity} is usually a term or a noun phrase in text that corresponds to a concept in the domain, \textit{relation} corresponds to a simplified sentence that is semantically complete and describes how entity1 and entity2 are connected, \textit{snippet} is a short portion of text from which the corresponding entity pair and the relationship is derived, and \textit{text\_anchor} is an HTML anchor that links the extracted tuple to the corresponding portion of the source document in the collection.
For example, from a paragraph on powers and responsibilities of a president the following tuple can be extracted: $<$president, parliament, ``President nominates the Cabinet members to the Parliament'', snippet, [URL][anchor]$>$.

We hand-tune the extracted tuples by correcting minor errors caused by the extraction of entities and relations.
These tuples are visualized as a knowledge graph where nodes are the entities and edges are the relationships between them. This visualization constitutes the lowest layer of the hierarchy. 

\subsubsection{Minimap Generation.}
The final component of this system generates a hierarchical representation of the search results by extracting a middle layer from the input Knowledge graph tuples and provides bidirectional mappings between all three layers. As noted earlier, we call this layer the minimap layer.

A natural result of the entity-relationship tuples extracted above is that some entities have a higher number of edges, i.e., are of higher degree. A higher edge count implies a larger number of connections to other entities in the graph; in other words, those entities with higher edge counts were more frequently linked with other entities in the document. We call these higher degree vertices, \emph{central concepts}, and hypothesize that one alternative to hierarchical faceted structures is to consider a multi-level view of a knowledge graph around central concepts. The multilevel view focusing on central concepts simply introduces information seekers to those entities or objects that are most frequently linked to other entities within the corpus. Generating the hierarchy becomes a thresholding task to appropriately scope the intermediate level of the visualization. Algorithm \ref{alg:ecc} describes this process more formally.

\begin{algorithm}[t]
    \caption{Extracting Central Concepts}
    \label{alg:ecc}
    \begin{algorithmic}[1]
        \Require $Nodes$: array of nodes in the knowledge graph, $min\_degree$: a pre-specified threshhold for the minimum degree of node to be considered as a central concept (starting value = 3), $max\_count$: an experimentally derived threshold for the maximum number of Central Concepts to be included in the middle layer (default value = 15).
        \Function{ExtractCC}{$Nodes, min\_degree, max\_count$}
        \While{true}
        	\State CentralNodes $\gets$ []
            \ForAll{$node$ in $Nodes$}
              \If{$node.degree \ge min\_degree$}
                  \State CentralNodes.add($node$)
              \EndIf
            \EndFor
        
          \If{$CentralNodes.size() \le max\_count$}
          \State \Return $CentralNodes$
          \EndIf
          
          \State $min\_degree$++
        \EndWhile
        \EndFunction
    \end{algorithmic}
\end{algorithm}

\subsection{Prototype Development}
\label{HKG_UI}
Given our hierarchical representation (DQ1), we must support mechanisms for viewing and interacting with the visualization (DQ2 and DQ3). 
In information retrieval, it is difficult to separate any visualization for representing search results from the interface that contains that visualization \cite{hearst06}. We iteratively designed an interface to support navigation of our hierarchical knowledge graphs via a series of pilot studies.

\begin{figure*}[htb!]
\centering
\includegraphics[width=\textwidth]{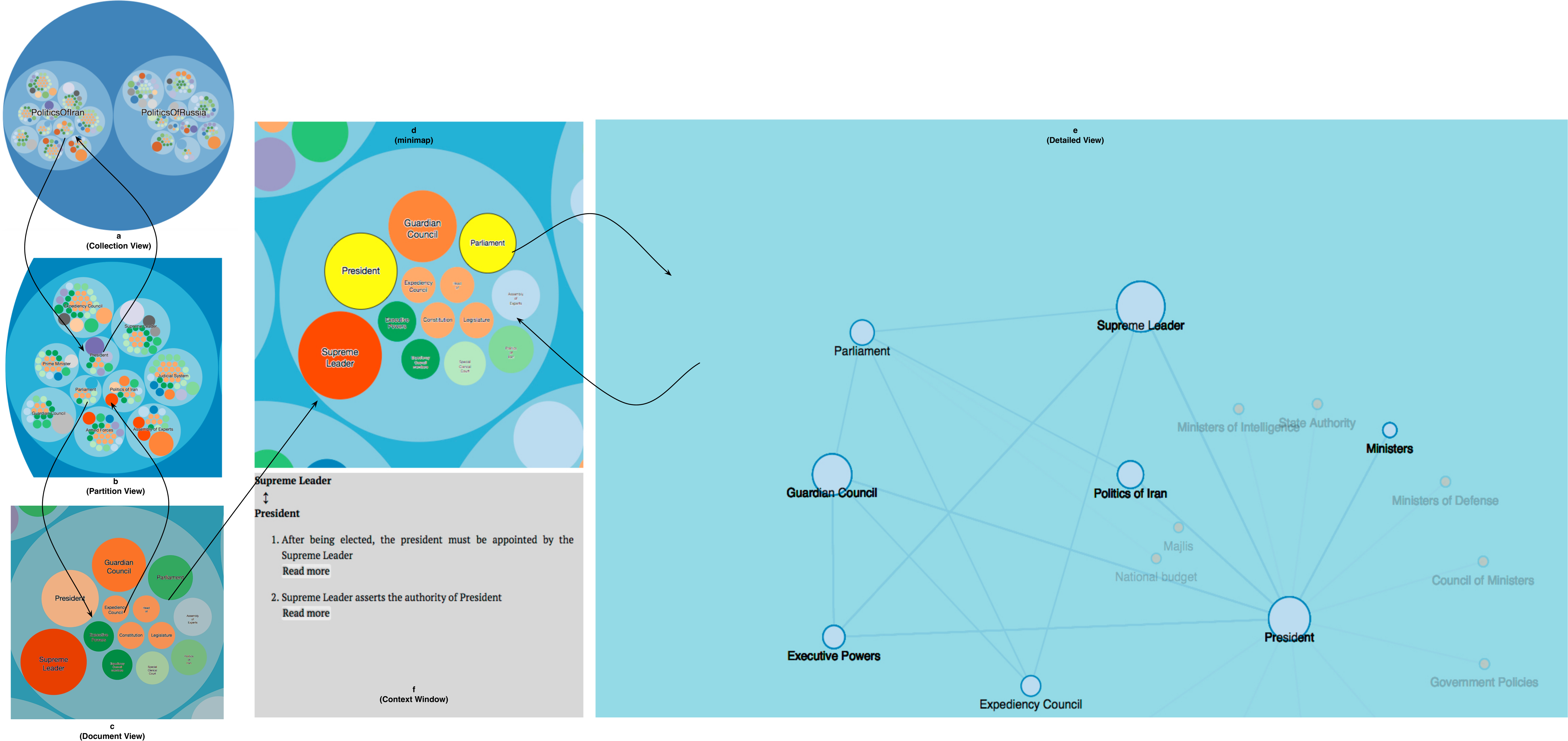}
\caption{Multi-layer Graph Interface: (a) Collection View; (b) Partition View; (c) Document View; (d) Minimap (i.e. Global View); (e) Detailed View (Local View); (f) Snippet Window}
\label{MLG_UI_Transitions}
\end{figure*}
Based on established literature and pilot studies we found that knowledge graphs can become overwhelming or confusing for participants \cite{cockburn96,komlodi07,olston03,sarrafzadeh16}. The overwhelming nature of the full knowledge graph leads to a need to create filtered views of our graph. These filtered views draw inspiration from the ``expand-from-known'' paradigm in information visualization \cite{van09}. Specifically, at the top level of the full corpus, a user selects a document, then a central concept from the minimap visualization. While preserving the entire knowledge graph, we alpha-blend all nodes in the knowledge graph \emph{except} those nodes directly related to the central concept from the minimap. Recall that the central concept is simply a high-degree vertex from the knowledge graph; therefore, the central concept and all its linked nodes are shown saturated. As a result, users can identify the central concept, linked entities, and can see closely related additional entities. Together, this focused detailed view seems to effectively support expand-from-known at the knowledge graph level.

As well, for the Hierarchical View, the biggest challenge to address was the disorientation among the participants during transitions between collection, minimap, and knowledge graph views, a common problem in interfaces that show multiple levels of abstraction. To address this disorientation (DQ3), we maintained the connection between the hierarchical view and the graph view in two ways. First, the user can move between the layers of Collection View and the Document View smoothly through a zooming functionality that changes the focus of the UI (see Figure \ref{MLG_UI_Transitions}).  Second, the interplay between the Document View and the Detailed View is designed such that the overview of the document is present at all times, in terms of a call-out on the left side of the screen, an actual \emph{minimap} as in computer gaming, which allows the user to maintain a sense of where he or she is while manipulating the fine-grained nodes and edges in the Detailed View.

The iterative process culminated in the final prototype shown in Figure \ref{MLG_UI_Transitions}.
In this interface, we see an initial overview, the Collection View that presents an overview of the underlying documents' structure in the collection (Figure \ref{MLG_UI_Transitions}-a).
The Collection View can potentially provide multiple partitions on the documents. Figure \ref{MLG_UI_Transitions}-b illustrates one partition of a collection.
As an information seeker drills down on each document, the view is altered (Figure  \ref{MLG_UI_Transitions}-c) such that an overview of the document is presented. The Document View provides a Global View of the corresponding document in terms of its central concepts. In this overview, the salient concepts in that article are visualized as circles of different sizes, where size indicates the frequency of occurrence in that article.  We used force and pack layouts (as part of the D3 library\footnote{http://d3js.org/}) to visualize the different layers of the knowledge graph representation.
 
The lowest layer of our representation is the Detailed View (Figure  \ref{MLG_UI_Transitions}-e). This view is a knowledge graph that represents entities and relationships between them.
The Detailed View, similar to our past work's graph interface \cite{sarrafzadeh16}, contains labeled nodes and unlabeled links between nodes. Nodes that represent entities with low frequency are hidden in the initial view, and only appear once a higher-frequency, connected node is clicked, ensuring that the graph does not become too cluttered. Once the user hovers over a node, that node and all connected nodes are highlighted, while the remainder of the graph is alpha-blended into the background. Clicking on a node can expand it by adding in its related nodes. Alternatively, clicking on a node can collapse its neighbours if they are expanded already. Nodes can also be dragged and placed at different parts of the canvas. This functionality can help with organizing the graph structure in a way that is more meaningful to the user and it can help with minimizing label overlap in the graph.
 
Edges can similarly be highlighted by hovering. By clicking on any edge, the user can see the relationship(s) between the two corresponding nodes (linked by the edge) in the context window located on the lower left side of the interface (Figure  \ref{MLG_UI_Transitions}-f). For each relationship in the context region, a hyperlink allows users to view the corresponding web page. 
\section{Evaluating Hierarchical Knowledge Graphs}
\label{HKG_Evaluation_PartI}
In our earlier work \cite{sarrafzadeh16}, the specific, complementary benefits of hierarchies and knowledge graphs were that hierarchies support a better global view of the search space, allowing participants to gain an appreciation of important topics whereas networks provide information on low-level entities and their relationship, thus reducing the need to read documents. The question explored in this section is whether this structure does, in practice, preserve these complementary benefits. We describe an experiment that specifically contrasts hierarchical knowledge graphs to participant behavior using hierarchies 
(based on underlying documents' table of contents)
and networks (uni-level knowledge graphs).  We analyze both quantitative data (gleaned from log files) and qualitative data (gleaned through observation of participant actions and post-experiment interviews with participants).

\subsection{Experimental Design}
\label{ExperimentalDesign_PartI}
To evaluate our HKGs' overall efficacy in supporting information seeking tasks as well as validating whether they can preserve the known strengths of network and hierarchical structures, we need to leverage a data set of extracted information, the representation of this information (i.e., our HKGs) and an interface to support browsing and sensemaking of this representation. We also need a set of control interfaces (i.e., reference interfaces that can be compared to HKGs).
As well, experimental design should replicate, as closely as possible, past work to ensure experimental validity. 

In recent work 
\cite{sarrafzadeh16}, we developed two interfaces for exploratory search: one knowledge graph interface and one hierarchical tree interface. To preserve experimental validity, we use identical interfaces as control interfaces. We also leverage the identical data sets, ensuring that topic is eliminated as a confound. Finally, we use exactly the same experimental task, ensuring that performance numbers are representative between experiments.

\subsubsection{Control Interfaces}

\begin{itemize}
\item
The first interface, a knowledge graph interface, functions as follows: As the interface starts, nodes that represent entities with low frequency are hidden in the initial view, and only appear once a higher-frequency, connected node is clicked. Users can also filter the knowledge graph by clicking on a node; when a user clicks on an edge, snippets and links associated with that edge are shown in a preview pane on the left side of the interface. 
\item
The second interface utilized a hierarchy (or a tree) structure to organize headings and sub-headings of the articles, as observed in each page's table-of-contents. When the user launches the application, the user is presented with a fully expanded tree. By clicking on any node within the tree, that portion of the Wikipedia document corresponding to the node is presented in the preview area at the left of the interface. 
\end{itemize}
\begin{figure}[ht!]
\centering
\includegraphics[width=0.70\textwidth]{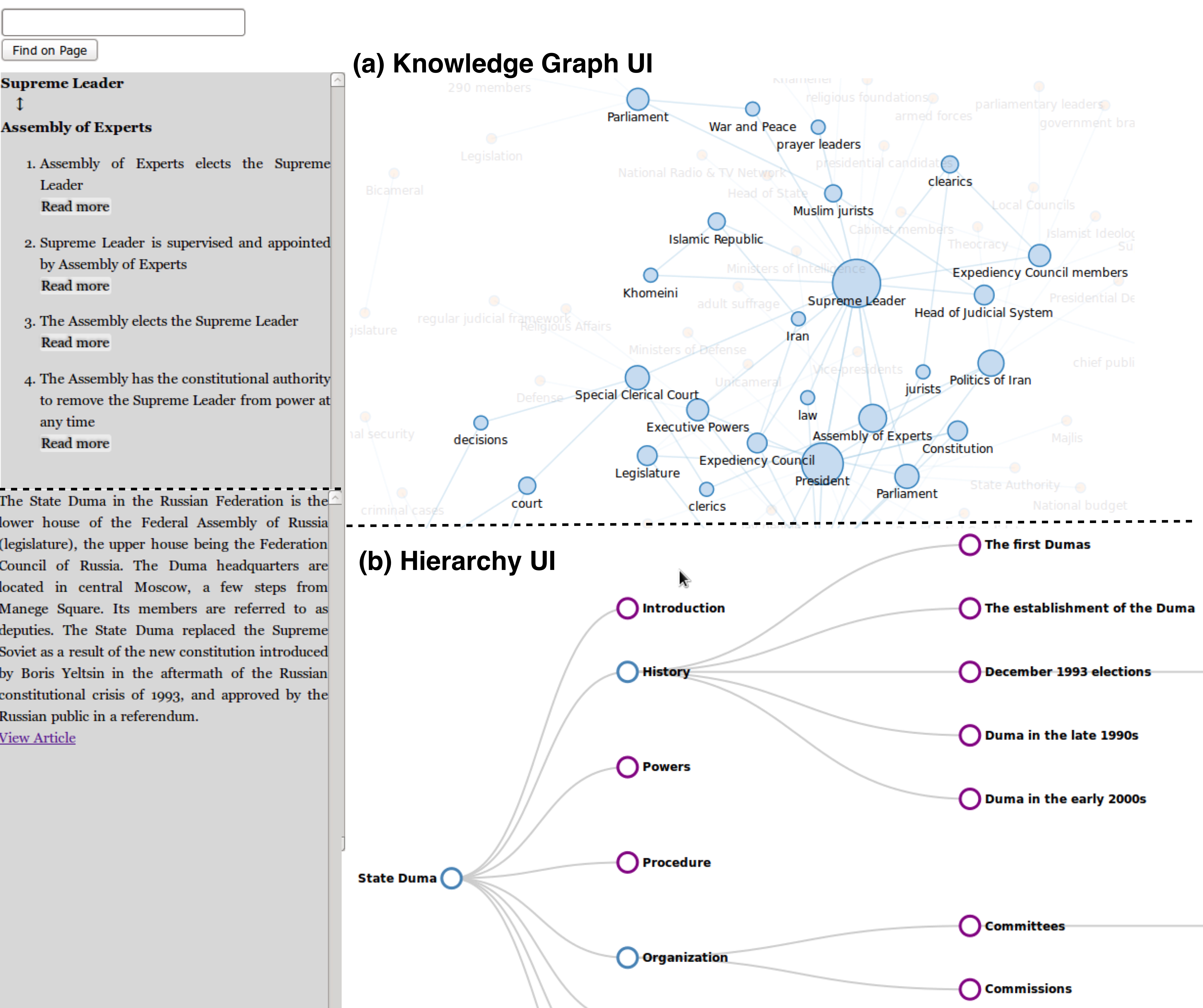}
\caption{Control interfaces for Knowledge graph and Hierarchy. More details in \ref{ControlUIs_details} and in \cite{sarrafzadeh16}.}
\label{controlinterfaces}
\end{figure}
Figure \ref{controlinterfaces} depicts these two interfaces. Contrasting these interfaces with Figure \ref{MLG_UI_Transitions} shows a similar preview pane for snippets. Links within the snippets function identically across all three interfaces. Section \ref{ControlUIs_details} elaborates on these interfaces and provide more Figures.

 
\subsubsection{Data Set}
\label{Dataset_Part1}
To populate our interactive applications, we created two distinct data sets: one focusing on history and the second on global politics. For the history data set, we focused on exploring former capital cities of Canada. For the politics search task, we created a data set representing governmental structures in Iran and Russia.

To create this data set, we first collected a set of Wikipedia articles by querying the Web using a popular search engine. We retrieved the top 10 articles in Wikipedia based on their relevance to three queries corresponding to three topics: ``Former Capital Cities of Canada,'' ``Political System of Iran,'' and ``Political System of Russia.''
To create our knowledge graphs from these data sets we leverage the Open Information Extraction system described in Section
\ref{KnowledgeGraphGeneration}.

\subsubsection{Search Tasks}
\label{SearchTasks_Part1}
Search tasks can be either simple (e.g., question answering) or complex (e.g., essay writing). With respect to the complexity level, each participant performed one Simple and one Complex task. We also used two different topics (i.e., History and Politics) to investigate the relation between the topic and content knowledge with the structure used to organize the retrieved information. The queries we asked people to find information to satisfy in our study were the following:
\begin{description}
\item[Simple Politics:] What governmental body or bodies are involved in the impeachment of the President of Iran and of Russia?
\item[Complex Politics:] Imagine you are a high school student who is going to write an essay on the Political Systems of Iran and Russia.
Knowing little about the presidents of these two countries, you wish to determine which president has more power. Find at least 3 arguments to justify your answer.
\item[Simple History:] As a result of which act were Upper and Lower Canada formed?
\item[Complex History:] Imagine you are a high school student who is going to write an essay on the History of Canada.
Knowing little about Canadian History, you wish to know which cities have served as a capital for Canada.
You would also like to understand the reasons behind moving the capital from one city to another.
\end{description}

To design these exploratory search tasks ~\cite{sarrafzadeh16}, we were guided by Marchionini's work on exploratory search~\cite{marchionini06}. These tasks combine aspects of knowledge acquisition/comparison (Marchionini's {\it learn} subcategory) with analysis, synthesis, and evaluation (Marchionini's {\it investigate} subcategory). In addition, the task descriptions closely follow Bystro\"m and Hansen's~\cite{bystrom05} recommendation that three levels of description should be used to specify a search task: a contextual description, a situational description and a topical description and query. 
Finally, in our prior work ~\cite{sarrafzadeh16}, we conducted quantitative and qualitative analysis on participants performing these exploratory search tasks and showed that these tasks were indeed complex (i.e., that they were ambiguous, open ended and exploratory in nature) and they were of sufficiently similar complexity as to limit topic effects. 

\subsubsection{Study Design}
 
Our study design was a 3 $\times$ 2 $\times$ 2 [interface, topic, complexity] mixed design.  For Knowledge graph and hierarchy, we leverage the data set from our past work \cite{sarrafzadeh16} in anonymized form. 
We add additional participants for our HKGs to yield our mixed design as follows.

For HKGs, each participant performed two different tasks, one simple and one complex.  The topic area (history or politics) differed for each of these tasks.  More formally, for these participants, our design was a 2~$\times$~2 full factorial mixed design, with topic and complexity as within subjects factors and complexity to topic assignment as a between subject factor. We counter-balanced the order in which the tasks were assigned to the participants.
 
Alongside the HKG participants, leveraging data from our past work \cite{sarrafzadeh16} adds two additional levels of Interface (hierarchical tree or knowledge graph) as a between subject factor.  Combining the data sets yield the 3 $\times$ 2 $\times$ 2 mixed design [interface, topic, complexity] with interface as a between subjects factor, and topic and complexity as within subjects factors.
 
\subsubsection{Participants}
In total we analyze data from forty seven participants. Twenty six participants, thirteen female, used hierarchies and knowledge graphs, the control interfaces. An additional twenty-one participants (4 female) used HKGs, the experimental condition, as a between subjects factor. All participants use the Internet on a regular basis to search for information.  Participants were aged between 18 and 45 years old (62\% were between 20 and 29 years old). Participants received a \$15 incentive for their participation.

\subsubsection{Procedure}

After introducing the study, participants were presented with an experimental interface (populated with an unrelated data set), and were given time to familiarize themselves with the interface and data structure.  Once participants had developed some comfort with the features of the interface ($\sim$ 3 minutes), participants completed a questionnaire assessing their familiarity with the topic used for the first task.  They were then given the description of their task (see above), and were asked to complete the task using the interface (15 minutes per task). Participants completed a post-task questionnaire that evaluated the experience; we used questionnaires provided by TREC-9 Interactive Searching track \footnote{www-nlpir.nist.gov/projects/t9i/qforms.html}  modified  to fit our experiment.  The same process was repeated for the second task. These questionnaires are provided in Appendix \ref{Appendix_Questionnaires}.
 
At the end of the second task, a semi-structured interview explored participants' experience using the interface. Interviews explored the conceptual usability of the visualization, the technical usability of the application and the efficacy of the interface for different types of search tasks.  Feedback on competing interfaces was also collected from participants. The details of these interviews are outlined in Appendix \ref{interviews}.
 
\subsubsection{Data Collection}

Alongside a mixed design of within subject and between subject factors, we perform a mixed methods analysis of both quantitative and qualitative data \cite{creswell2013research}. Data was captured as follows:

\textbf{(a)} The interface was instrumented with a logger which monitored movement on the computer screen and participants' interactions with the system. Interactions collected included node or edge clicks, snippets read, articles viewed, and time spent reading the articles. In HKGs, the transition between the layers and switches between the MiniMap and the knowledge graph were captured.

\textbf{(b)} Two assessors evaluated the quality of answers provided by the participants for each of the search tasks independently. Simple queries were rated as either correct or incorrect.  Complex questions were rated on a scale. Scores for all queries were normalized to reflect a value in the range [0, 1]. Inter-assessor reliability was evaluated using Pearson coefficient and an overall value of 0.97 for simple queries and 0.94 for complex queries was found. Appendix \ref{interviews} describes the marking scheme used for assessing the quality of essays provided for complex tasks. 

\textbf{(c)} We captured field notes during participant interactions, audio recorded all sessions, transcribed final interviews, and collected questionnaire data.  This data was analyzed collectively using open coding to extract low-level themes and axial coding to identify thematic connections between elements.  Coding was performed incrementally as each participant's data was collected, and saturation was found after coding qualitative data from field notes and transcripts for 15 of our 21 participants.

\subsubsection{Hypotheses and Research Questions}
\label{hypotheses_part1}
Quantitative data allows us to test the following hypotheses:
\begin{itemize}
\item
Hierarchical knowledge graphs result in fewer document views and less time spent reading documents than do hierarchical trees.
\item
Hierarchical knowledge graphs exhibit statistically similar behaviors to Knowledge Graphs.
\end{itemize}

Alongside hypothesis testing, our log data provides insight into whether hierarchies are used in hierarchical knowledge graphs and on whether task complexity affects the use of hierarchies. As well, to triangulate quantitative data, we leverage our qualitative data to compare and contrast the nature of the hierarchies between the tree interface and the hierarchical knowledge graphs and to understand whether the hierarchies provide similar affordances.  In the remainder of this section, we present, first, our quantitative analysis of logged date, followed by our analysis of qualitative and observational data captured during our study.  We conclude by discussing the implications of our results. 

\subsection{Results: Quantitative Analysis}
\label{Results_Quantitative_PartI}
Below, we present a quantitative assessment of our HKGs assuming perfect IE algorithms are in place.
In particular, we use data collected from search logs to test our main hypotheses stated in Section \ref{hypotheses_part1}.

Scoring of participant responses by independent evaluators and log file analysis produced the quantitative measures in Table \ref{table1} for Hierarchical Knowledge Graphs (H. Graphs), Hierarchical Trees (H. Trees), and Knowledge Graphs (K. Graphs). Rows represent measures for Marks (MK), Nodes clicked (NK), Edges Clicked (EC), Document Views (V) and Document View Time (VT). We break each measurement out by two query levels, Simple and Complex, as described previously.


\begin{table}[ht]
    \begin{tabular}{cc|c|c|c|}
        \cline{3-4}
        & & \cellcolor[gray]{0.8} H. Graphs & \cellcolor[gray]{0.8} H. Trees & \cellcolor[gray]{0.8} K. Graphs \\ \hline
        \multicolumn{1}{|c}{\multirow{5}{*}{\rotatebox[origin=c]{90}{Simple}}} &
        \multicolumn{1}{|l|}{\cellcolor[gray]{0.8} MK}                         & 0.43 (0.21)   & 0.32 (0.20)     & 0.37 (0.14)     \\ \cline{2-5}
        \multicolumn{1}{|c}{} & \multicolumn{1}{|l|}{\cellcolor[gray]{0.8} NC} & 11.4 (8.6)    & 19.0 (10.04)    & 11.38 (9.4)    \\ \cline{2-5}
        \multicolumn{1}{|c}{} & \multicolumn{1}{|l|}{\cellcolor[gray]{0.8} EC} & 18.3 (8.9)    & NA              & 27.15 (12.9)              \\ \cline{2-5}
        \multicolumn{1}{|c}{} & \multicolumn{1}{|l|}{\cellcolor[gray]{0.8} V}  & \textbf{2.38 (1.61)}   & \textbf{6.08 (2.49)}     & \textbf{2.38 (3.00)}     \\ \cline{2-5}
        \multicolumn{1}{|c}{} & \multicolumn{1}{|l|}{\cellcolor[gray]{0.8} VT} & \textbf{145.6 (153.7)} & \textbf{1430.9 (2302.8)} & \textbf{211.6 (228.0)} \\ \hline

        \multicolumn{1}{|c}{\multirow{5}{*}{\rotatebox[origin=c]{90}{Complex}}} &
        \multicolumn{1}{|l|}{\cellcolor[gray]{0.8} MK}                         & 0.62 (0.18)  & 0.57 (0.28)      & 0.58 (0.16)      \\ \cline{2-5}
        \multicolumn{1}{|c}{} & \multicolumn{1}{|l|}{\cellcolor[gray]{0.8} NC} & 13.38 (9.2)   & 20.09 (17.7)     & 26.23 (19.12)     \\ \cline{2-5}
        \multicolumn{1}{|c}{} & \multicolumn{1}{|l|}{\cellcolor[gray]{0.8} EC} & 23.09 (12.7)  & NA               & 41.07 (19.4)               \\ \cline{2-5}
        \multicolumn{1}{|c}{} & \multicolumn{1}{|l|}{\cellcolor[gray]{0.8} V}  & \textbf{2.15 (2.13)}  & \textbf{4.38 (2.24)}      & \textbf{4.38 (2.24)}      \\ \cline{2-5}
        \multicolumn{1}{|c}{} & \multicolumn{1}{|l|}{\cellcolor[gray]{0.8} VT} & \textbf{103.4 (97.6)} & \textbf{985.38 (1848.3)} & \textbf{78.76 (131.5)} \\ \hline
    \end{tabular}
    \caption{Hierarchical (H.) Graphs vs. Hierarchical Trees and Knowledge (K.) Graphs: Mean (Standard Deviation) values for marks (MK - average independent evaluator scores), clicks on nodes (NC) and edges (EC), document views (V), and document view time (VT). Bolded dependent variables exhibited significant differences in post-hoc testing.}
    \label{table1}
\end{table}




\subsubsection{Hypotheses Testing}
Multivariate analysis of variance with respect to interface (tree versus graph versus hierarchical graph), topic (history versus politics), and task (simple versus complex) for Marks (MK), Views (V), and View Time (VT) shows a statistically significant effect of interface  ($F_{6, 172} = 7.126, p < 0.001, \eta^2 = 0.2$) and task ($F_{3, 86} = 12.22, p < 0.001, \eta^2 = 0.3$) on dependent variables. Post-hoc factor analysis using Tukey correction indicates that the tree interface exhibited statistically significantly higher numbers of document views than both hierarchical graphs and knowledge graphs. As well, the tree exhibited statistically longer reading times than hierarchical graphs ($p<0.05$), but not than knowledge graphs (p = 0.064) in our analysis.  Hierarchical graphs and knowledge graphs did not differ significantly in their effects on any dependent variables. Task significantly impacted the marks but no other variables.

Clicks are not directly comparable between H.Trees, H.Graphs, and K.Graphs, as edges are not clickable in hierarchies (NA value in Table 1). Performing pairwise comparison between H.Graphs and K.Graphs, our analysis showed no statistically significant effect on dependent variables ($F_{3, 30} = 0.752, p > 0.5, \eta^2 = 0.70$), including node click and edge click behavior.

Given the above analyses, we reject both null hypotheses and conclude that our hypotheses are supported by our data set. Hierarchical Knowledge Graphs preserve the advantages of Knowledge graphs over hierarchical trees in both reading time and in document views. Focusing specifically on our hierarchical graph, we find that our hierarchical graph has statistically lower document views (61\% fewer document views, on average) and time reading (90\% less time reading documents) than does hierarchical trees and that its behavior is statistically indistinguishable from the prior observations of knowledge graph interfaces.  Furthermore, the effect size measures, $\eta^2$, are significantly above the threshold (0.14) typically considered to be a large effect, lending support to these differences being sufficiently large to be meaningful. In summary, our quantitative results support our hypothesis that our hierarchical knowledge graphs fully preserve the quantitative advantages identified by our prior work \cite{sarrafzadeh16} for knowledge graphs over hierarchies.



\subsubsection{Additional Quantitative Analysis}
Given the statistically indistinguishable nature of HKGs and Knowledge Graphs, one question is if (and whether) intermediate hierarchical representations are used. It is possible that Hierarchical Knowledge Graphs are indistinguisable from Knowledge Graphs because users ignore the hierarchy and simply leverage the knowledge graph.

\begin{table}[ht]
\begin{center}
    \begin{tabular}{ l|c|c|c|}
    \cline{2-4}
    & \cellcolor[gray]{0.8} GlobalView & \cellcolor[gray]{0.8} MiniMap & \cellcolor[gray]{0.8} DetailedView \\ \hline
    \multicolumn{1}{|l|}{\cellcolor[gray]{0.8} Simple Task} & 27.03\% & 14.61\% & 58.0\% \\ \hline
    \multicolumn{1}{|l|}{\cellcolor[gray]{0.8} Complex Task} & 23.83\% & 17.24\% & 58.90\%  \\ \hline
    \end{tabular}
    \caption{Percentage of Time spent on each of Global View, Minimap and Detailed View}.
\label{table2}
\end{center}
\end{table}

To specifically explore this question, we looked at how much time users spent on each of the provided views in our HKG interface. 
Overall, our data indicated that participants took advantage of all three layers relatively similarly across both Simple and Complex tasks. Further, while the the time spent on detailed view dominates other views (58\% for the simple task and 59\% for the complex task), over 40\% of time was spent on additional views in the hierarchy (Table \ref{table2}). Looking specifically at how participants spent their time in different layers of the hierarchy (i.e., utilizing different views of the data) for different tasks we see that the time spent at the detailed view is similar for both levels of complexity. On the other hand, participants seem to spend less time in MiniMap than Global for the simple task (Pairwise t-tests with Tukey correction yields statistical significance, $p < 0.01$). For Complex task, however, time in Global versus mid-level are not statistically different ($p > 0.1$). Essentially, in the complex task, sensemaking is split between global and minimap views of the hierarchy more equitably, i.e., the minimap is particularly useful during our complex tasks.

\begin{figure}[ht!]
\centering
\includegraphics[width=0.9\columnwidth]{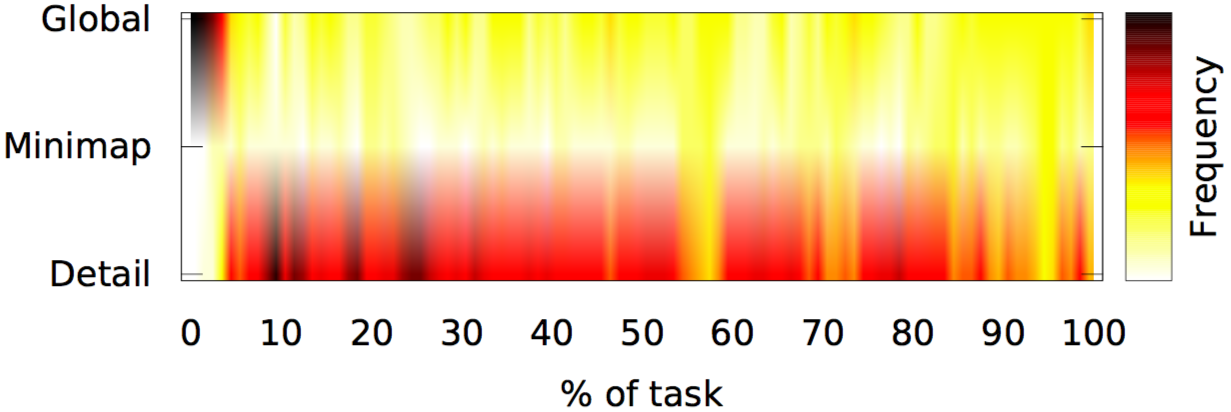}
\caption{Heatmap visualizing the patterns of users navigating views in HKG for intervals of 1\% of task length.}
\label{heatmap}
\vspace{-2mm}
\end{figure}

We also explored usage patterns of views. Figure \ref{heatmap} is a heatmap that visualizes use of different views for intervals of 1\% of task length. Early in the task, we see frequent use of the global view. While difficult to see, MiniMap usage peaks just after the halfway point in the task, but there is no strong concentration of use. The hierarchy, and particularly the MiniMap visualization, seems to be used throughout the task.







\subsection{Results: Qualitative Analysis}
\label{Results_Qualitative_PartI}
The next analysis we perform involves participant perspectives on hierarchical knowledge graphs as a representation of search results. We were particularly interested in the overviews knowledge graphs provide for the information space and their contrast with table-of-contents-based hierarchies.

To address these questions, we performed open-coding of observations, transcripts, and questionnaire data. We coded incrementally, and saturation occurred after fifteen participants were coded. We coded all participants for completeness. Once open coding was complete, axial coding and thematic analysis was performed collaboratively by the researchers.
We present three themes arising from our qualitative data analysis: Supporting Exploratory Search Tasks, Imposing a Structure versus Open Exploration and the Self-Orienting nature of HKGs.
\vspace{-0.4mm}
\subsubsection{Supporting Exploratory Search Tasks}
As noted in our study design, we incorporate two exploratory information seeking tasks with different levels of complexity. In post-experiment interviews the participants were able to compare how different task complexities are supported by the assigned interface.

The hierarchical graph representation was found to provide more support for the Complex Task (i.e., more open ended and exploratory tasks such as essay writing or learning) versus Simple tasks (such as question answering and specific knowledge finding).
This observation seems to be true for any multi-level structure which provides an overview and allows a gradual immersion into details: Finding a specific piece of information to satisfy a simple query is best done using a traditional search engine. 



Looking specifically at HKGs and complex tasks, the overview allowed participants to identify the central concepts of a domain at a glance and the size of the circles indicates their prominence in the corresponding article.
As many participants noted, `relevance' or `prominence' of a concept with respect to the main topic or the domain they are exploring is an important asset in Complex search tasks. This qualitative observation may explain the more equitable use of the MiniMap representation for complex search tasks noted in our quantitative analysis. Complex tasks required synthesizing, rationalizing, and comparing, which seem to require more awareness of the entire data set.

This identification of central concepts was also linked to a perception of value of the MiniMap as a starting or entry point into the topic of the document being examined. Several participants articulated a belief that the overview provided by central concepts helped with ``going from knowing nothing to having a plan,'' ``learning terminology,'' ``relevance, importance, or prominence,'' and ``objectively learning about a domain.'' In particular, the \emph{objective} nature of central concepts was cited by many participants as key to their utility.

As White and Roth \cite{white09} point out, exploratory search is motivated by complex information problems, poor understanding of terminology and information space structure, and often a ``desire to learn.'' Vakkari \cite{vakkari2000} also argues ``more support is needed in the initial stages of a task,'' when users have an unstructured mental model.
Inspired by Kim \cite{kim99}, in our prior work \cite{sarrafzadeh16} we found that hierarchical trees provide this benefit in unfamiliar domains. A strength of our design of hierarchical knowledge graphs is that it enables the user to engage in two alternative navigation paradigms. Users can exploit overview layers to explore the collection at a higher level followed by targeted immersion in the detailed view.

\subsubsection{Imposing a Structure versus Open Exploration}
While most participants were unanimous that the hierarchical representation imposes a [subjective], [rigid] structure onto the information space, their attitude towards this phenomenon varied. The level of domain knowledge and the complexity of the search tasks were found to be the major factors affecting their attitude.

When the searcher is dealing with a domain where he has limited knowledge, they are more open to accepting the structure that the representation imposes. Both hierarchical trees and hierarchical knowledge graphs incorporate imposed structures.
Participants articulated a variety of advantages to structures: it was ``easier to follow,'' ``contained important aspects'' that ``simplified focus,'' and guided participants in ``where to go'' or ``what steps to follow.''
With respect to hierarchical trees, some participants simply ``trusted'' the designer of the hierarchy (e.g., the author of an article) to be ``logical'' or ``rational'' in the way he broke down things.
This was particularly true for participants with limited knowledge of a topic domain and replicates findings by our prior work \cite{sarrafzadeh16} and Amadieu et al. \cite{amadieu10} that low knowledge learners benefited from hierarchical structures in free recall performance and exhibited reduced disorientation. 

In the case of higher domain knowledge, our participants were split in their preferences and attitudes.
Some still trusted the logic behind the layout of a hierarchical trees and the fact that their knowledge of the domain can guide them to find what they want using this hierarchy. They trusted the designer to place items in close proximity to where the item should be. Other participants strongly opposed the rigid structure of a hierarchy, feeling it was ``not the way I think,'' ``based on the mindset of the author,'' or ``did not match the domain structure.''

One interesting perspective of the multi-layer graph representation which presents central concepts of a domain as an overview for each document is that it reflects the knowledge graph concepts. This reflection made it, for many participants, more flexible and exploratory, a window into the knowledge graph. Many participants commented on this phenomenon, noting it was ``guiding but not imposing,'' ``more open,'' ``sparked interest'' in the lower level structure, or was ``visually appealing'' and ``fun.''
\vspace{-1mm}
\subsubsection{Self-Orienting or Relative Positioning}

One main advantage of the Hierarchical Tree visualization in our prior work \cite{sarrafzadeh16} was the explicit connections between nodes (categories or headings) in the representation. These edges help in two ways:
\begin{enumerate}
\item At a glance, you can tell why a concept appeared in this overview, or in this domain. To whit, the hierarchical structure exists the way it does because of a human author's decision.
\item The Path from the root to each of these nodes in the Tree Layout can provide useful information on where a concept is positioned relative to the topic.
\end{enumerate}

We \cite{sarrafzadeh16} previously noted that participants may perceive a domain to have a derivative/hierarchical structure or a multi-faceted structure.
If salient relationships are viewed as derivative or hierarchical (e.g., `is-a' relationships), then a tree can best capture this view of data; whereas, if salient relationships are more heterogeneous and resist structure as a hierarchy, that disadvantages the hierarchies.

This is not the case in our MiniMap, where the connection between each of these main concepts and the main topic is unknown at first glance. Central concepts are simply extracted based on their high connectivity with other concepts within a specific document within a corpus. However, it is also true that it would be quite surprising if highly linked concepts were not, somehow, important components of any individual document. The more pervasively they link, the more they interconnect with other concepts, the more important it is to understand them and their relationship. In this way, HKGs become self-orienting for our participants.

\subsection{Discussion}
The primary goal of this experiment was to explore whether we could combine benefits from both knowledge graphs and hierarchies into one data structure for visualizing search results.  Examining Table \ref{table1}, we note that our HKGs significantly reduce documents read and reading time as compared to hierarchical trees; qualitatively, the results are almost identical to results obtained for traditional knowledge graphs. This provides support for both hypotheses: that HKGs reduce the need to read documents compared to hierarchies generated from tables of contents, and that HKGs perform, both statistically and qualitatively, on par with knowledge graphs.
Via log analysis, we also provide  evidence that the hierarchy is used by participants (see Figure \ref{heatmap}).

One important consideration is whether our control interfaces and our HKG's interface introduce confounds (i.e. is the HKG interface introducing quantitative advantages to HKGs independent of any representation advantages). Obviously, one cannot test every conceivable interface and prove that one has an optimal interface given an infinite number of possible interface variants:  instead, we wish to determine whether control interfaces represent consistent principles of behavior versus the experimental interface containing our HKGs.  Our results provide some support to the premise that our control interfaces (knowledge graphs and hierarchies) represent effective control conditions.  First, the form of the two control interfaces are highly similar, with differences primarily a function of data structure \cite{sarrafzadeh17}, yielding some confidence that they are mutually similar.  Second, recall that the goal of HKGs was to preserve the benefits of Knowledge Graphs and incorporate some of the advantages of hierarchies.  Considering Table \ref{table1}, we can see that HKGs and Knowledge Graphs have highly similar user behaviors versus metrics captured in logs, an indication that our interface results in similar user behaviors on the knowledge graph structure that is the basic representation contained within these interfaces.  

Qualitative data from our participants indicate that hierarchies grounded in tables-of-contents are more familiar, easier to follow, and more focused. 
This is primarily because the tree layout explicitly represents connections between nodes, which helps with understanding how and where a concept fits in a bigger picture. 
This in turn helps users  orient themselves in the data. The author-vetted nature of hierarchical tables-of-contents was also perceived to be an asset absent from our hierarchical knowledge graphs.  The hierarchies in our knowledge graph were viewed slightly differently, as a more data-driven representation, which gives them a certain cachet with respect to the unbiased nature of topic selection but does violate the ordered structure of tables-of-contents inspired hierarchies.

Alongside a contrast between the hierarchical structures in trees and in our hierarchical knowledge graphs, it is also useful to contrast certain aspects of behavior for hierarchical knowledge graphs (HKG) and knowledge graphs. The goal of hierarchies in HKG was to help users self-orient within the data, to develop an overview of the data. This was one identified benefit of the hierarchical view provided by the tree interface. Looking to our quantitative data to contrast our two graph-based structures, we see a reduction in the number of nodes clicked in hierarchical knowledge graphs versus knowledge graphs, and, in univariate analysis this reduction is statistically significant for complex tasks. We believe that this may be because hierarchical knowledge graphs allow users to self-orient within the data, to develop an impression of the overall significant topics. However, because we did not qualitatively evaluate within subjects the differences between hierarchical knowledge graphs and knowledge graphs, we leave further analysis for future work.

In terms of limitations, one concern with our analysis of HKG behavior revolves around the ecological validity of our results, particularly in light of hand tuning of information extraction. As we noted in our experimental design section, we used automated algorithms to generate knowledge graphs \cite{KG4ES_tech} and extracted hierarchies from tables-of-contents or headings within documents. However, we then hand tuned both the hierarchies (adding low-level sectioning to documents) and information extraction results (refinement of co-referencing). We do note that there are benefits to hand tuning.  Specifically, to ensure that confounds are \emph{not} present in our results, hand-tuning (or at least manual verification) is essential; otherwise, error-prone algorithms and poorly structured data could influence the effectiveness of any individual representation of search results, focusing the data around the algorithmic failures as opposed to the nature of hierarchies versus graphs.  
However, it is also the case that any realistic assessment of the potential of representations should also consider its resilience to realistic errors in underlying algorithms, and it is this question we explore in the following section.


\section{Impact of Information Extraction Errors on HKGs}
\label{HKG_Evaluation_PartII}
\begin{figure}
    \centering
    \includegraphics[width=0.9\columnwidth]{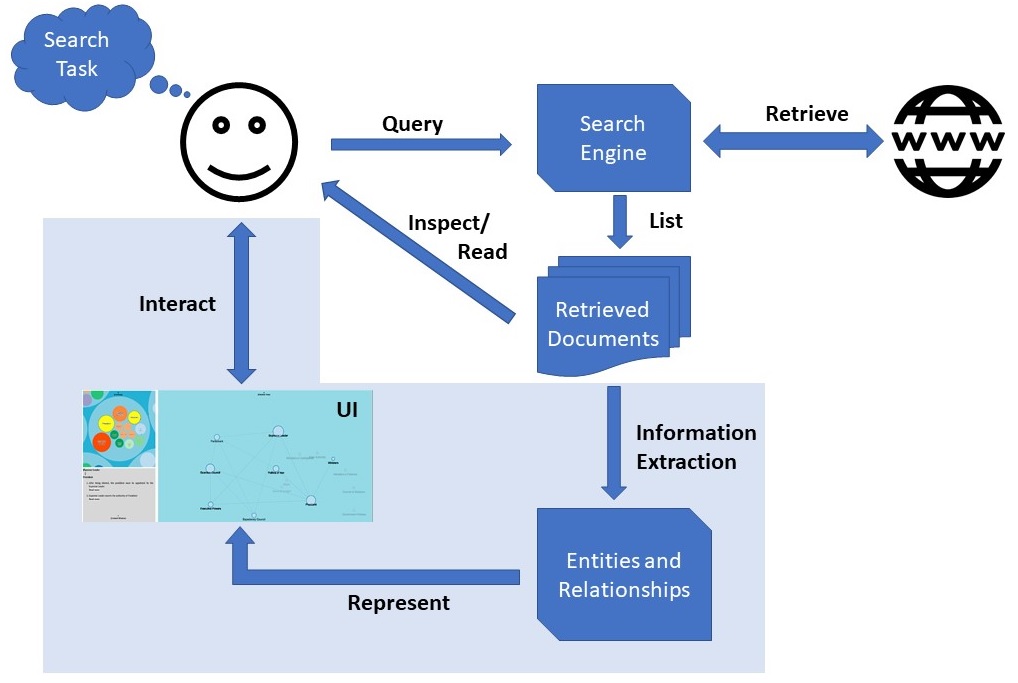}
    \caption{In traditional search interfaces, a user issues a query, inspects the retrieved documents, and, if necessary, reformulates the query to refine the retrieved documents until the retrieved set of documents satisfies their search task. In our exploratory search system, shaded in light blue, Information Extraction creates a set of Tuples which are then represented in a UI. Through interaction with this UI, exploratory search tasks are supported.}
    \label{fig:IEErrors}
\end{figure}

In this section, we evaluate the performance of HKGs in light of errors in information extraction. To understand why we wish to explore the impact of errors in information extraction, consider Figure \ref{fig:IEErrors}. In typical web search, users formulate queries, inspect retrieved documents, and either view documents or, if they find that the returned documents are not exactly appropriate, reformulate queries to refine the set of documents retrieved. Because a user can directly examine the results of a query retrieval operation, the user can refine the search query to modify the retrieved documents as needed. However, when performing information extraction, one challenge that the user faces is a limited ability to influence the quality of extracted information. Even if the set of retrieved documents is correct, errors in information extraction propagate through the representation of the entity-relationship tuples.

Any realistic system that incorporates representations such as HKGs to support exploratory search is presented with two options: either ensure that information extraction is perfect -- a constraint that is beyond the present ability of natural language processing algorithms -- or perform its task in the presence of errors (i.e., deficiencies in either precision or recall).  Given that information extraction is not yet a solved problem, in this section we explore the behavior of HKGs given uncorrected entity-relationship tuples generated by an information extraction system. Accordingly, these uncorrected tuples result in either missing information (i.e., lower recall) or incorrect/non-relevant information (i.e., lower precision).  To detail our experimental design, we start with instantiating a system to support exploratory search which follows the structure described in Figure \ref{fig:IEErrors}. Next, we describe the study design, our participants and the experimental procedure. Finally, we describe the data we capture from each participant.

\subsection{Experimental Design}
\label{PartII_ExperimentalDesign}

To examine the impact that different levels of precision and recall have on exploratory search while using HKGs, we use two different information extraction outputs.  One set represents the raw, uncorrected output of an IE algorithm; the second represents human-corrected output, used in the previous section to evaluate the potential of HKGs. 
We use these two outputs to populate our hierarchical knowledge graphs and leverage the interface that we designed (described in Section \ref{HKG}) to support interaction with these HKGs. The following subsections describe different aspects of our experimental design in more detail.



\begin{figure*}
\centering
\includegraphics[width=0.95\textwidth]{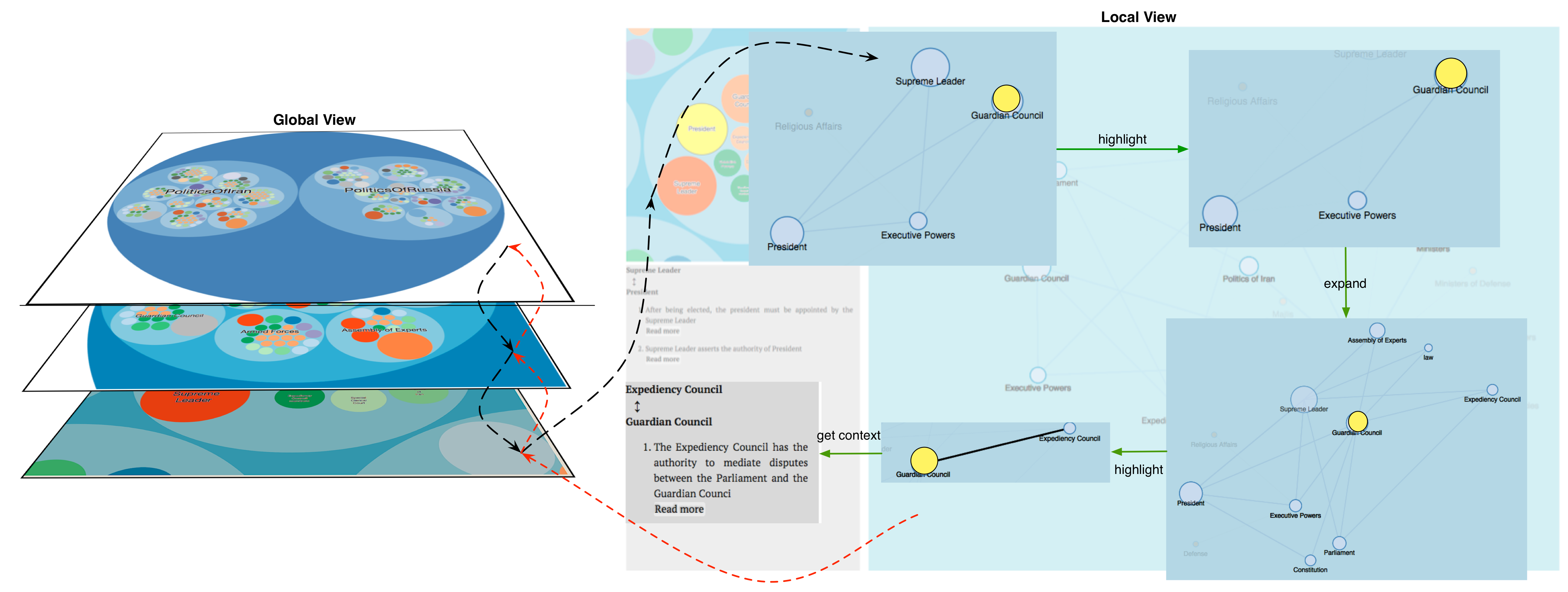}
\caption{The populated exploratory search interface leveraged in our study. This interface, represents search results as a hierarchical knowledge graph structure which enables the user to engage in two alternative navigation paradigms. While users can frequently exploit the overview layers to explore the collection at a higher level followed by targeted immersion in the detailed view, they can alternatively descend within an area of interest in a graph and continue exploring in the detailed view itself by manipulating nodes and edges and getting more context on relationships.}
\label{fig:interface}
\end{figure*}

\subsubsection{Automatic Information Extraction}
\label{AutomaticIE}

For our experiment, we follow the same procedure described in Section \ref{Dataset_Part1} to generate knowledge graphs.
The resulting entities and relations are emitted along with the snippet, from which the entities and relation were selected, and a HTML anchor to the snippet in the source text.



\begin{figure}[htb!]
\centering
\includegraphics[width=0.5\textwidth]{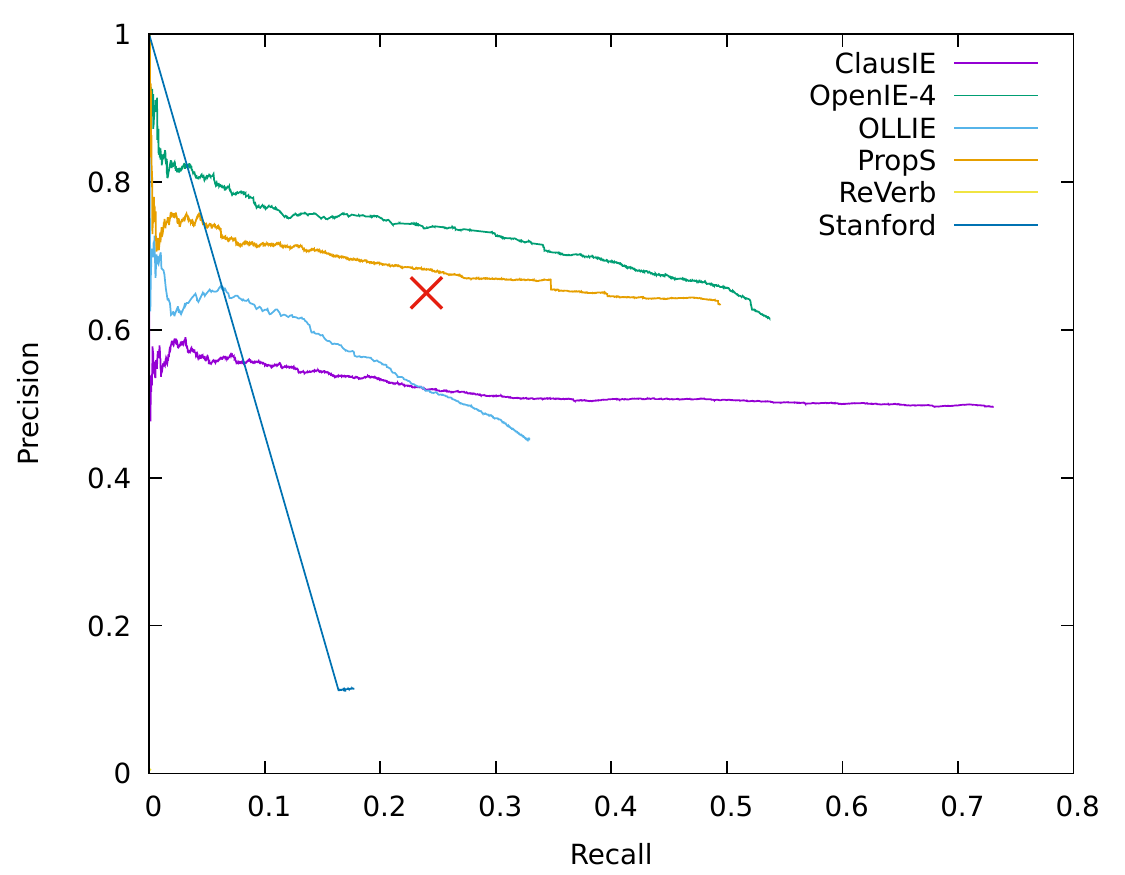}
\caption{Precision-recall curve for the different Open IE systems using Stanovsky et al.'s toolkit~\cite{stanovsky16}. The X represents our system.}
\label{PRcurves}
\end{figure}

To determine the performance of our IE system, we used the Open-IE benchmarking toolkit~\cite{stanovsky16}. As can be seen in Figure \ref{PRcurves}, our system performance is representative of the breadth in state-of-the art systems; specifically, we have tuned relative precision and recall of our system such that it achieves a precision of 0.65 and recall of 0.24 for the task, which is the mid-point of the precision and recall curves presented in the toolkit~\cite{stanovsky16}. 

Some may question our decision to choose approximately median (as opposed to optimal) performance for our information extraction system.  If our goal was to investigate the best possible performance of systems leveraging automatic information extraction, choosing the best system would be justifiable. However, our goal is to examine the effect that errors in information extraction have on performance. Accordingly, selecting the top performing system would yield a biased experiment which would be limited to insights about the best-performing algorithms, whereas more representative performance across a class of algorithms allows us to generalize to systems of varying qualities. That is, such systems should perform at least as well as our system relative to manually tuned extractions.  
\vspace{-2mm}


\subsubsection{Hierarchical Knowledge Graphs}

As we saw in Section \ref{HKG_Evaluation_PartI}, HKGs can combine aspects of both hierarchical structures with the ability to explore spatially the relationships between information. 
While the results discussed in Sections \ref{Results_Quantitative_PartI} and \ref{Results_Qualitative_PartI} provided evidence for the efficacy of HKGs in preserving the known strengths of networks and hierarchies,
we relied on perfect information extraction to determine these benefits. To accurately assess how HKGs fare in ``real-world'' scenarios, it is necessary to compare usage of HKGs under less than ideal circumstances (i.e., error-prone information extraction).


\subsubsection{Exploratory Search Interface}
As our experiments require users to complete exploratory search tasks (outlined in the following section), we required an interface to support interaction with HKGs. To maintain consistency and to control for confounds, we re-use the interface developed for the experiments of Section \ref{HKG_Evaluation_PartI}. While better interfaces for HKGs may exist, we do not believe that widely varying the interface will give us an accurate picture of user behaviour. 
Figure \ref{fig:interface} shows a alternate view of this UI which enables two alternative interaction paradigms to explore the search results.
As described in Section \ref{HKG_UI}, this interface allows smooth transition between overview and detailed views of HKGs (supporting overview-filter-detail-on-demand search \cite{shneiderman96} and expand-from-known searching \cite{van09}). 

One advantage of leveraging an existing interface is that we eliminate the confound of interface effects on user behaviour.  Specifically, because our interface is identical to past work, we can compare human-corrected output to output from past work.  Then, if the output is similar, any deviations in results between different error levels are attributable to the effect of extraction errors rather than interface idiosyncrasies.

\subsubsection{Exploratory Search Tasks}
\label{SearchTasks}

Again, to preserve consistency with past results and to control for confounds, we leverage two complex exploratory search tasks described in Section \ref{SearchTasks_Part1}.
As a reminder, we provide the task descriptions as used in this study:
\begin{description}
\small
\item[Complex Politics:] Imagine you are a high school student who is going to write an essay on the Political Systems of Iran and Russia. 
Knowing little about the presidents of these two countries, you wish to determine which president has more power. Find at least 3 arguments to justify your answer.
\item[Complex History:] Imagine you are a high school student who is going to write an essay on the History of Canada. 
Knowing little about Canadian History, you wish to know which cities have served as a capital for Canada. 
You would also like to understand the reasons behind moving the capital from one city to another.
\end{description}

\subsubsection{Acquiring Knowledge Graphs at Different Levels of Quality}
As noted earlier in this section, we use two different information extraction outputs to populate our HKGs. One set represents the raw, uncorrected output of an IE algorithm; the second represents human-corrected output.
These expert crafted tuples form what can be considered the best possible execution of the extraction algorithm and represent a ``best effort" for knowledge graph creation and so form our {\it gold standard}. Accordingly, automatically running the extraction algorithm should yield an inferior graph that contains errors that would impact user performance. 

\paragraph{\textbf{Characterizing Precision and Recall of Automatic vs Hand-Tuned IE}}

To test whether precision and recall rates differ between {\it Auto}matic and {\it Gold} graphs, we note that the {\it Gold} generated graph for the History task contains 2,957 entity-relationship tuples and the Politics graph contains 3,231 tuples. In contrast, the {\it Auto}matically generated graphs contain 1,782 and 2,735 tuples, respectively.  
We ran the automatic graph through the Open-IE benchmarking toolkit with the manually curated graph as ground truth. For Politics, the automatic graph achieved a Precision of 0.56 and a Recall of 0.33. For History, the automatic graph achieved a Precision of 0.7 and a Recall of 0.31. These results are in line with what we might expect given our earlier results from running our IE algorithm on the Open-IE benchmark (See Section \ref{AutomaticIE}). 

\subsubsection{Study Design}

Different components of our evaluation framework were discussed in previous subsections. As a final step,
we  describe a controlled in-lab experiment to investigate the effect of an error-prone Open IE system's output (i.e., the automatically ran system) on users conducting exploratory search tasks. The main goal of our study is to investigate the effect of errors on user behaviour. To facilitate this, our study design was a 2 x 2 [error-level, topic] mixed design with error-level and topic as within-subject factors and error-level to topic assignment as a between-subject factor. The two levels of error were \textit{Gold} and \textit{Auto}, which correspond to the manual and  automatic run of the algorithm previously described. The topics correspond to the History and Politics tasks. The order of error-level to topic assignment is fully counter balanced to mitigate any order or learning effect. This resulted in a full-factorial design with 4 groups. Participants were randomly assigned to groups.

We recruited 25 (11 female) participants from different areas of Science, Arts and Business, Math and Engineering for this study, all of whom use the Internet on a regular basis to search for information. Participants were aged between 20 and 45 years old (80\% were between 20 and 29 years old). They received a \$15 incentive for their participation.

\subsubsection{Procedure}



Our study began with a brief introduction and a short familiarization period ($\sim3$ minutes) with the experimental interface that was populated with a completely unrelated dataset. The goal of this familiarization period was to allow the participants to ``figure out'' the nuances of the search interface unconstrained by a particular task. Following this period, participants were given an initial questionnaire intended to gauge their prior knowledge of their first task. The questionnaire combined a self-assessment of their own prior knowledge and a list of three questions, ordered by increasing difficulty, to provide an objective assessment.

Following this questionnaire, participants were presented with the description of the task and were asked to complete the task using the interface (15 minutes per task).
To capture data on participants' use of the interface, they were asked to provide reference sentences that indicate the information included in the essays was not purely based on their prior knowledge of the topic. 
After 15 minutes had elapsed, participants completed a post-task questionnaire that evaluated their experience of the task. We used questionnaires provided by TREC-9 Interactive Searching track~\footnote{Available at \url{www.nlpir.nist.gov/projects/t9i/qforms.html}.} modified to fit our experiment. Additionally, we inquired about each participant's perceived assessment of the quality of the knowledge graphs in terms of the information they provided, including whether they noticed errors, inconsistencies, or missing information. The same process was repeated for the second task.
Appendix \ref{StudyResources} provides more details on the questionnaires used during the study ($\sim$ \ref{Appendix_Questionnaires}) and the assessment of participants performance in completing the complex search tasks ($\sim$ \ref{MarkingSchemes}).

Following the completion of second task, a semi-structured interview was conducted to explore participants' experience of the interface. This interview focused on the conceptual usability and efficacy of the interface for different tasks as manifested in a participant's perception of task complexity, the strategies they employed to complete tasks, and their belief as to whether this type of interface supports these types of tasks. We also inquired about their perceptions of knowledge graph quality, the information they found using the graph, and whether they noted any differences in graph quality between the tasks.

As a final step, participants were presented with a list of factoid questions on both topics to collect feedback on how their behaviour may change when performing question answering style tasks. Participants were given a small amount of additional time to familiarize themselves with the interface to complete this type of task. Once they were comfortable, they were asked to find the answer to one final question that was erroneous as a means to determine whether they would notice such an error during their information seeking process. We collected strategies and changes in quality perception when using the interface for question answering tasks (see Appendix \ref{interviews} for more details on interviews).


\subsubsection{Behavioural Data Collection}
Following Creswell~\cite{creswell2013research}, we collect data to perform a mixed methods analysis of both quantitative and qualitative factors. We log all participant interactions with the system, which includes the node or edge clicked, articles viewed, time spent reading articles, as well as more topological data (e.g., interactions with the UI itself, such as transitioning between layers or using the ``minimap''). In addition, we captured field notes and audio during all participant sessions, transcribed the final interviews, and collated the questionnaire data. Quantitative data was analyzed using SPSS.  Qualitative data was analysed using open coding to extract low-level themes and axial coding to identify thematic connections. This process was conducted incrementally as each participant completed the study and we attained qualitative saturation after 14 out of 25 participants. Sampling continued to ensure that participant set cardinality was sufficient to support statistical discrimination.

As participants were required to provide essays for each search task they completed, we had two independent assessors evaluate the quality of these answers. To ensure consistency with prior work, we reused the marking scheme designed in ~\cite{sarrafzadeh16}. To aid in further analyses we normalized all scores to be in the range [0,1] and inter-assessor reliability was found to be 0.90 using the Pearson coefficient (i.e., reasonable agreement).

\subsubsection{Hypotheses and Research Questions} \label{Hypotheses_PartII}
Quantitative data allows us to test the following hypotheses:
\begin{itemize}
    \item Automatically generated hierarchical knowledge graphs result in a lower performance (i.e. task outcomes - measured by essay qualities) than do manually curated HKGs.
    \item Automatically generated hierarchical knowledge graphs result in more document views and more time spent reading documents (i.e. proxies for effort) than do manually curated HKGs.
\end{itemize}

Alongside hypothesis testing, our log data provides insight into whether other factors such as the type of task, rather than the error condition, affect the task outcomes.
As well, to triangulate quantitative data, we leverage our qualitative data to first characterize how the two types of exploratory search tasks were perceived by the participants and next explore whether and how the quality differences between our \textit{auto} and \textit{gold} HKGS impacted participants behaviors and perceptions during exploratory search tasks. 
\subsection{Results: Quantitative Analysis}
\label{Results_Quantitative_PartII}

Because past work suggests that searchers are able to adapt their behavior to the system performance by controlling their effort (e.g., issuing more queries \cite{smith08}, reading more documents \cite{turpin01}, allowing more interaction with the search results \cite{smucker10}) in the search process, our interaction logs can be used as a proxy for the effort the searcher is contributing during the search task and whether there are differences when comparing Auto and Gold conditions. Of particular interest to us was clicking on edges to read the relationship label (EdgeClick), viewing snippets to get the context around extracted relation label (Snippet), viewing articles to continue the exploration using the articles themselves (ViewArticle), and overall time on task (Duration). 

\newcommand\mcx[1]{\multicolumn{1}{|c|}{\cellcolor[gray]{0.8} \textbf{#1}}}
\newcommand\mcy[1]{\multicolumn{2}{|c|}{\cellcolor[gray]{0.8} \textbf{#1}}}

\begin{table*}[bhtp]
\begin{center}
\resizebox{\textwidth}{!}{
    \begin{tabular}{ c|c|c|c|c|c|c|c|c| }
    \hhline{~|--------|}

    &
    \mcy{Group 1 (HA - PG)} &
    \mcy{Group 2 (HG - PA)} &
    \mcy{Group 3 (PG - HA)} &
    \mcy{Group 4 (PA - HG)} \\
    \hhline{~|--------|}
    
    &
    \cellcolor[gray]{0.9} Auto &
    \cellcolor[gray]{0.9} Gold &
    \cellcolor[gray]{0.9} Auto &
    \cellcolor[gray]{0.9} Gold &
    \cellcolor[gray]{0.9} Auto &
    \cellcolor[gray]{0.9} Gold &
    \cellcolor[gray]{0.9} Auto &
    \cellcolor[gray]{0.9} Gold \\
    \hline

    \mcx{Mark} &
    0.52 (0.12) & 0.56 (0.14) &
    0.72 (0.15) & 0.66 (0.15) &
    0.40 (0.16) & 0.65 (0.11) &
    0.65 (0.11) & 0.64 (0.16) \\
    \hline

    \mcx{Snippet} &
    12.71 (6.7) & 16.30 (12.30) &
    11.50 (5.80) & 11.83 (6.77) &
    14.83 (3.97) & 10.0 (2.83) &
    12.0 (8.55) & 13.67 (5.72) \\
    \hline

    \mcx{ViewArticle} &
    3.0 (2.08) & 4.0 (3.21) &
    0.67 (0.82) & 1.5 (1.38) &
    3.17 (1.72) & 1.0 (1.56) &
    1.67 (1.50) & 1.5 (1.87) \\
    \hline

    \mcx{Duration} &
    169.57 (120.76) & 189.58 (174.87) &
    53.83 (58.36) & 19.33 (22.85) &
    163.33 (90.81) & 44.0 (63.34) &
    69.0 (81.03) & 78.17 (95.80) \\
    \hline

    \mcx{EdgeClicks} &
    20.0 (10.20) & 27.86 (27.51) &
    25.5 (11.04) & 21.83 (11.50) &
    27.0 (14.08) & 27.67 (15.02) &
    30.5 (15.04) & 21.67 (10.33) \\
    \hline
    \hline

    \cellcolor[gray]{0.8} &
    History & Politics &
    History & Politics &
    History & Politics &
    History & Politics \\
    \cline{2-9}

    \multirow{-2}{*}{\cellcolor[gray]{0.8} \textbf{Prior Knowledge}} &
    1.0 (0.1) & 0.14 (0.38)&
    1.0 (0.89) & 0.5 (0.83) &
    0.67 (0.82) & 0.0 (0.0) &
    1.3 (0.82) & 0.83 (0.98) \\
    \hline

    \end{tabular}
    }
    \caption{Contrasting relative performance and effort effects: Mean (Standard Deviation) values for dependent variables. Each group indicates the assignment of automatic or gold extraction (A or G) as well as the order of the Search Tasks (H for History and P for Politics).}
    \label{effort_performance_table}
\end{center}
\end{table*}

Table \ref{effort_performance_table} summarizes means (standard deviations) for these quantitative measures as well as the scoring of participant responses by independent evaluators (Mark) and estimates of their prior knowledge for each topic.

\subsubsection{Data Validity Tests and Sample Size Calculations}

An important first question, before we analyze the collected data to test our main hypotheses specified in Section \ref{Hypotheses_PartII}, is whether user behavior is consistent between our two studies.  To examine this question, we contrast the distribution of measurements between this study and the previous study to ensure the newly collected data is consistent with the previous results. For this analysis we contrast only the data collected for the \textit{complex task condition} in the first study with the data collected for the \textit{gold condition} in the second study. 
We performed a two (independent) sample Kolmogorov-Smirnov test for three main dependent variables ViewArticle, Duration and Marks between the two studies. The results showed that these measurements are highly likely to be sampled from the same distribution for ViewArticle ($D(20, 25) = 0.400, p > 0.95$), Duration ($D(20, 25) = 0.500, p > 0.95$) and Marks ($D(20, 25) = 1.000, p > 0.2$).

Next, for each of our measures, we performed a sample size power estimate \cite{howell2009} to ensure that our sample size is sufficient to identify statistically significant differences in dependent variables.
We used means, standard deviations and t-statistics at 0.975 in our power estimate to support a two-tailed analysis of effects at the 95\% confidence interval. We found that that sample sizes of approximately 12 to 18 were sufficient to analyze Mark, ViewArticle and Duration.  We then applied a Shaprio-Wilkes test of normality and found that Marks are normally distributed, but that ViewArticle and Duration exhibit modest right skew. We note that this skew is also obvious by inspection from Table \ref{effort_performance_table} by examining means and standard deviations.  To accommodate for non-normal distribution in ViewArticle and Duration, we calculated the coefficient of variation of the distribution (standard deviation divided by mean). Considering ViewArticle and Duration, this value yields coefficients between 0.7 and 1.25, a modest skew.  Modelling this via the central limit theorem, sample sizes of greater than 20 are sufficient to obtain normal mean sample values within 20\% of the mean; we set our sample size to 25 to ensure sufficient statistical power for pairwise statistical tests on ordinal data. \footnote{See https://statisticsbyjim.com/basics/central-limit-theorem/ for modeling moderately skewed data.}

\subsubsection{Statistical Contrasts}

We performed a repeated measures ANOVA with ErrorLevel (Auto versus Gold) as a within subject effect and Group as a random factor. The group variable encodes error level to topic assignment as well as the ordering of the tasks. For example, in Group 1 (HA - PG), participants performed the \textbf{H}istory task with \textbf{A}utomatically extracted information first and \textbf{P}olitics task with \textbf{G}old data next.
Dependent variables were Mark, ViewArticle, Duration, Snippet and EdgeClick.

\begin{figure*}[bht]
  \centering
  \subcaptionbox{Marks\label{fig:marks}}{
    \includegraphics[width=0.31\textwidth]{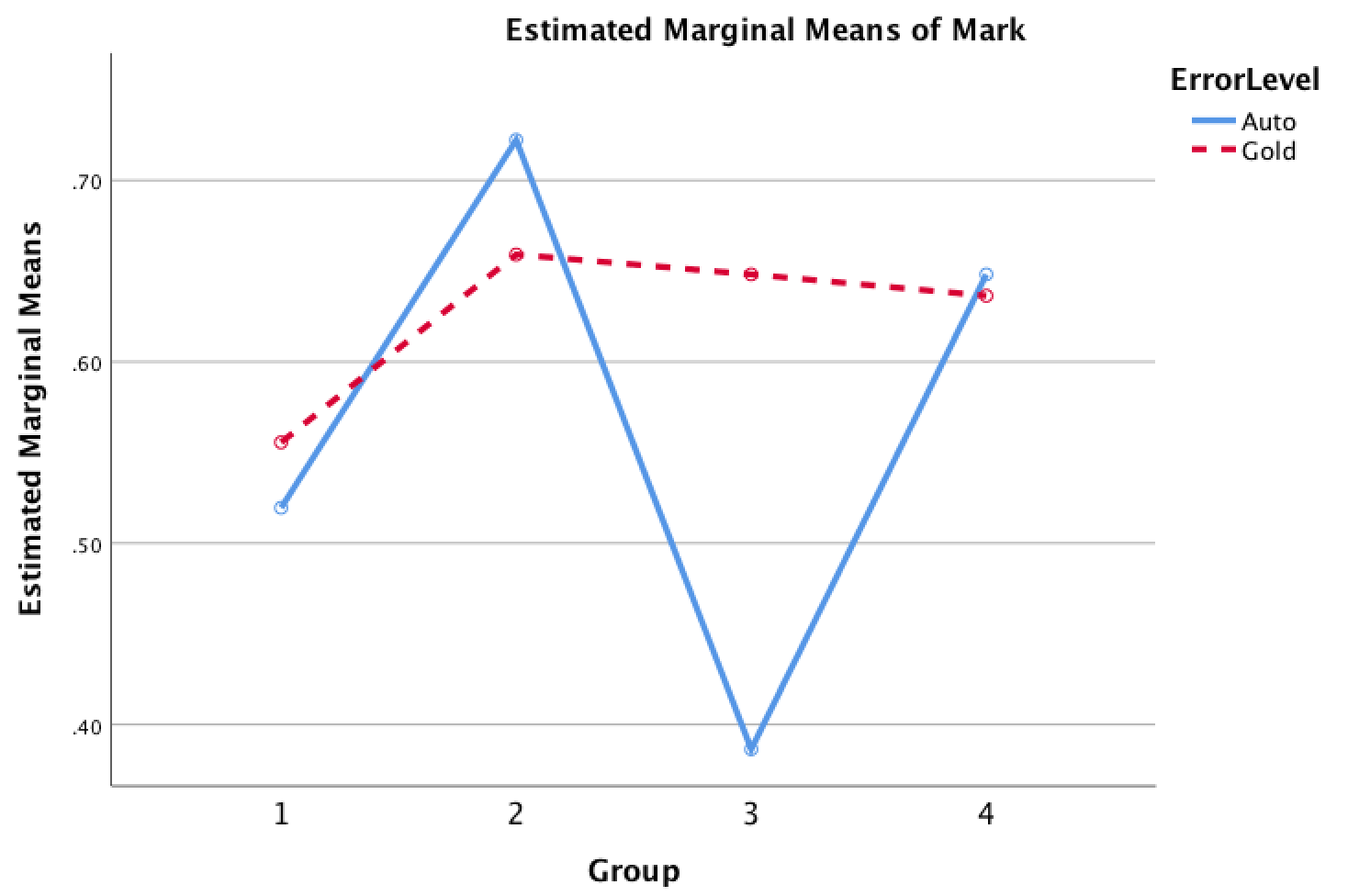}
  }
  \subcaptionbox{ViewArticle\label{fig:va}}{
    \includegraphics[width=0.31\textwidth]{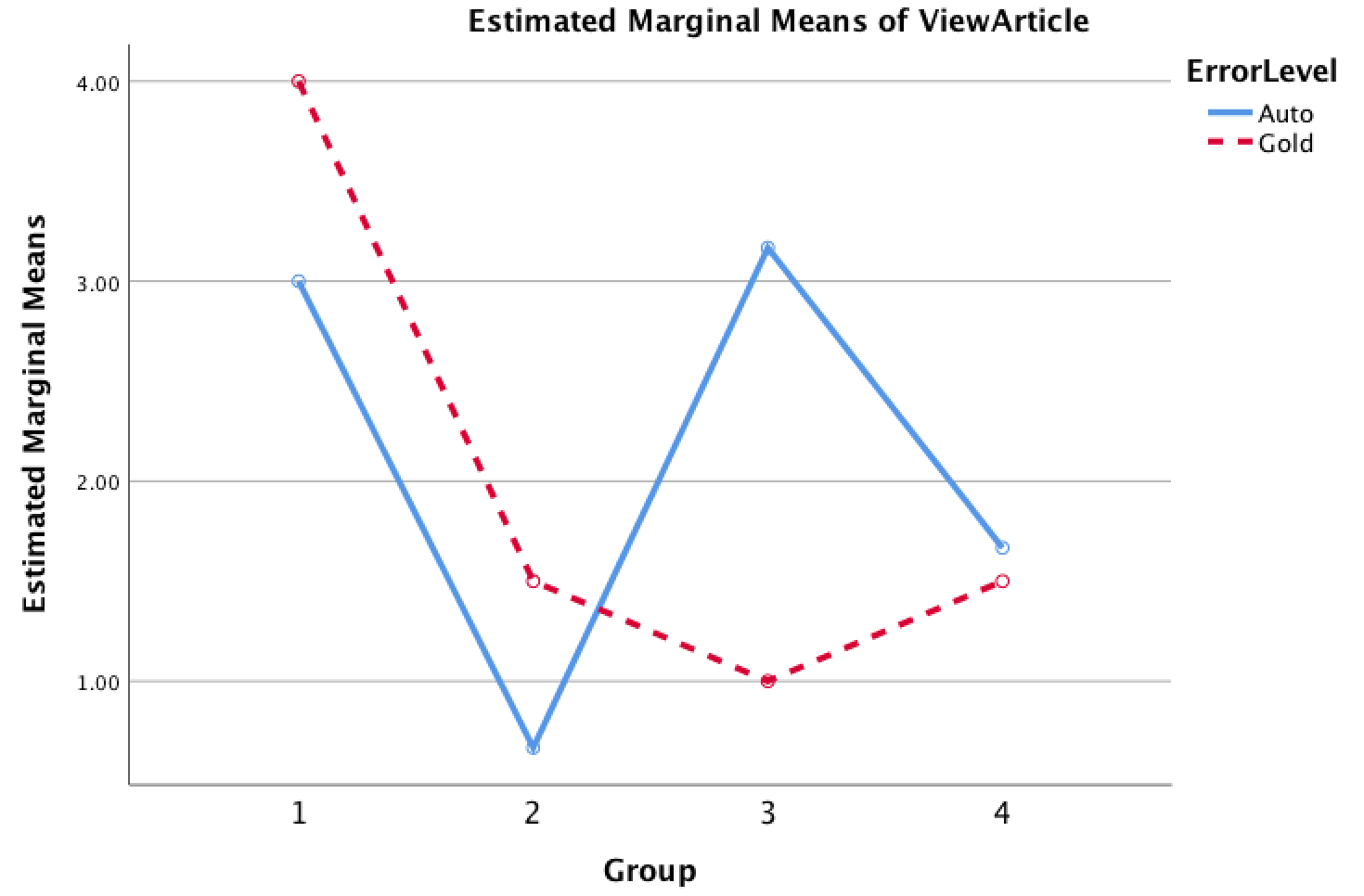}
  }
  \subcaptionbox{Duration\label{fig:duration}}{
    \includegraphics[width=0.31\textwidth]{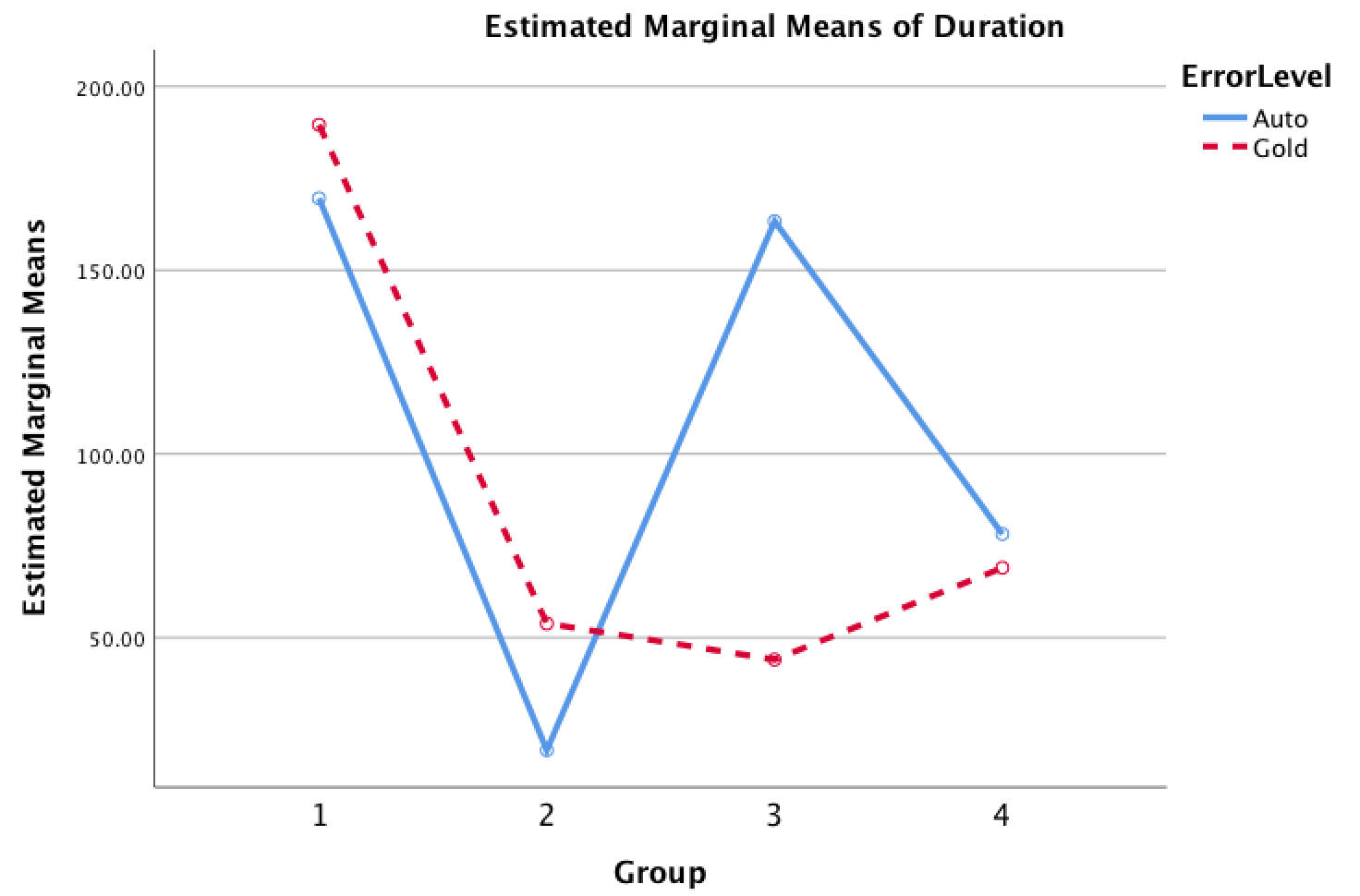}
  }
  \caption{Effects of Error Level and Group on (a) Marks, (b) ViewArticle and (c) Duration.}
  \label{fig:Impact_on_DVs}
\end{figure*}

Overall RM-ANOVA indicated that there was no statistically significant effect of ErrorLevel on Mark ($p > 0.1$), as a measure of search performance, nor was there any interaction between Group and ErrorLevel. 
There was, however, a significant effect of the group as a random factor on Mark ($F_{3, 25} = 4.808, p < 0.05, \eta^2 = 0.3$) and Duration ($F_{3, 25} = 3.935, p < 0.05, \eta^2 = 0.4$).
Post-hoc analysis, using Tukey correction, indicates that Group 1 and 3 had lower marks than Group 2; Group 4 was in the middle, not significantly different than any other group. As well, participants in Group 1 spent more time reading articles than participants in Group 2.
In terms of view article, there was no significant effect of error level, but there was a significant Group x ErrorLevel interaction ($F_{3, 25} = 4.269, p < 0.05, \eta^2 = 0.4$). 
Figure \ref{fig:Impact_on_DVs} demonstrates these trends in our data.

\subsection{Results: Qualitative Findings}
\label{Results_Qualitative_PartII}
In order to triangulate quantitative data, we leverage our qualitative data to first understand how the designed exploratory search tasks were perceived by the participants and next dive deeper on characterizing exploratory search behavior with error prone knowledge graphs. The analysis of our qualitative interviews and think-alouds indicated that participants varied in their perception of complexity of the tasks while there were reasonable level of agreement on different attributes of each search task.
Furthermore, different ways searchers can be impacted by error-prone knowledge graphs and the factors that help or hinder noticing these errors while engaged in complex information seeking activities were identified.
Finally, we contrasted different types of search tasks at different levels of complexity and specificity of the information being sought and how they differ in the way they are impacted by errors in retrieved information.
The following subsections elaborate on these themes. 
\subsubsection{Characterizing the Designed Exploratory Search Tasks}
The qualitative analysis of our post task interviews resulted in a rich characterization of the two exploratory search tasks and how they compare with respect to their perceived complexity as well as main distinguishing attributes. We found that our participants were split in their perception of the complexity of these two tasks and that a variety of factors such as prior knowledge of the searcher, interpretation of the task, and the strategies employed impacted their perception. 

Participants who found the History task more difficult ([P1-5, P7, P9, P10, P15, P20]) mentioned a variety of reasons including \textit{``representation does not match the perceived domain structure''} [P3, P10, P15], \textit{``Politics is a more familiar domain''} [P5, P7, P10], \textit{``needed a deeper exploration''} [P9], \textit{``searching without a keyword''}[P3], \textit{``more detailed fact finding''}[P6, P8, P13].

A different group of participants, however, perceived the Politics tasks to be more difficult [P4, P6, P11-13, P17-18, P21-22, P24-25]. The most common rationales included \textit{``Politics task asks a higher level question and needs a higher level understanding of a domain''} [P8, P10], \textit{``involves learning, complex reasoning and interpretation''} [P6, P8, P11, P13], \textit{``is multi-faceted''} [P13], \textit{``involves objectively learning about a domain that requires more time''} [P11, P18] and \textit{``more unfamiliar and technical terminology''} [P13].

Furthermore, these two search tasks, although both exploratory, were perceived to possess different attributes. Overall, the Politics task was found to be more subjective, required more high level information and belonged to the Learn category of exploratory search tasks \cite{marchionini06}. This task was also found to be a precision-oriented task as exploration was done mostly at a higher level and learning a few accurate facts about how each president functions locally or globally could suffice for comparison.
The History task, on the other hand, was seen as more objective and was characterized as more of a finding task that involves more detailed information and fact retrieval, closer to the Investigate category of exploratory search task \cite{marchionini06}. This task also seemed to share attributes of recall-oriented tasks, which requires a higher number of relevant facts being retrieved to satisfy the search task.
\vspace{-1.6mm}
\begin{myquote}{0.2in}
    \textit{``For the Politics task I thought that I had to learn something, whereas for the other one I didn't feel I need to learn anything. 
    Like in order to make any reasonable statement I will have to actually learn something about these countries.''} [P8]
\end{myquote}

One interesting distinguishing attribute was the notion of \textit{subjectivity} that was incorporated into the Politics task description. All participants were unanimous in the observation that the Politics task, as opposed to the History task, is asking about a subjective topic, \textit{the power of a president}, which can be interpreted and rationalized differently. This in turn impacted the perceived complexity of this task based on how participants defined power and whether or not they perceived this task as an ``objective learning task'' or a ``subjective goal-oriented search''.
We observed that participants who started the task without any predetermined outcome of the task (e.g., which president is more powerful) found the task more difficult as they first opted to objectively learn about the Politics of both countries in order to identify the aspects on which compare the presidents as well as find out which president is indeed more powerful. As well, to our surprise, the majority of the participants who found the Politics task more difficult had indeed more prior knowledge of this topic (i.e., were familiar with the politics of at least one of the two countries). One potential reason for this observation was that participants with higher prior knowledge of the topic were aware of more facets that can define power and they had higher expectations of what constitutes a reliable comparison between the two presidents. On the other hand, participants who interpreted this task as a goal-oriented search with the mere objective of finding evidence that conforms to the president that they believed had more power, were more satisfied with their search outcome and as a result perceived the task to be easier to perform.

\subsubsection{Perceived Quality of Generated Knowledge Graphs}
The second theme that arose from our qualitative analysis explored how knowledge graphs generated at different levels of quality, as measured by precision and recall, were perceived by participants performing exploratory search tasks. 
In our post-task interviews, the participants engaged in an in-depth discussion of how the knowledge graphs and the information they presented were perceived in terms of their quality, accuracy and whether or not they had a good coverage of the information they were expected to represent.
Given that the automatically generated knowledge graphs had a recall as low as 0.32 and a precision of 0.60 (averaged over Politics and History topics), we expected this difference to be noticeable by the participants. 

To our surprise, participants did not notice any significant quality differences between the knowledge graphs they interacted with across the two search tasks and they were not able to identify the automatically generated ones when were specifically asked to do so.
Participants, when asked specifically about the quality of the information represented by the graphs and whether or not they noticed errors, inconsistencies, incomplete or unreadable sentences, missing information, and so on, they mentioned that they found the graphs to be comparable across both tasks.
While a few participants mentioned they noticed minor errors such as typos, or duplicates they didn't think graphs associated with one of the tasks were significantly better or worse than the other. 

Next, participants were told that one of their tasks did use automatically generated knowledge graphs while the other used expert-curated data and were asked to guess which task was done with the automatically generated graphs. The majority of participants were still not able to identify the task with lower quality graphs. We received some comments regarding the information that they expected to see or the structure they expected the graphs to be represented with but differed from the actual experience.
\begin{myquote}{0.2in}
    \textit{``I think if you come across contradictory knowledge when you're not confident about your prior knowledge then I would maybe try to accept both of them and try to justify them in my mind.''} [P5]
\end{myquote}

\subsubsection{Factors impacting the recognition of errors}
The final theme from our qualitative interviews explored the factors that may help or hinder noticing of errors in automatically extracted information and contrasted the impact of these errors between lookup and exploratory search tasks. We identified two categories of factors that can influence the likelihood of noticing errors during information seeking activities: (a) Individual Differences; (b) Task Complexity and Scope. We elaborate on these two factors next.

\paragraph{\textbf{Individual Differences or User Effects}}
The interview continued with discussing the reasons that might have hindered participants from noticing errors or lower quality of graphs for one of the tasks.
We found that the intent of the user performing the task (i.e., information seeking versus validating information), their confidence in the task (i.e., their prior knowledge of the topic as well as their language competency) and different types of cognitive biases impact the likelihood of noticing errors.
\begin{myquote}{0.2in}
    \textit{``Now I can see that mostly when I'm reading sentences I look at keywords only. Especially when it's not written in my first language. And sometimes it feels like all English sentences should make sense, which is not the case in my first language, Persian. I do validate sentences as I read them in my first language!''}[P13]\\
    \textit{``I  think there are multiple factors to consider. One thing is about my confidence in some language. When I read sentences in English, I don't question the validity as often as I do when I read text in my first language. 
    So I'm more focused on understanding what the sentence is trying to say and not so much on spotting potential grammar issues, typos or inconsistent information. [...]
    And for more technical texts it's a mix of unfamiliarity and lack of confidence in my own knowledge of the domain as well as a higher level of trust in the validity of information that makes me accept what I read as facts.''}[P13]
\end{myquote}

\begin{myquote}{0.2in}
    \textit{``I think if you come across contradictory knowledge when you're not confident about your prior knowledge then I would maybe try to accept both of them and try to justify them in my mind.
    E.g. See ``President is the highest authority in Iran'' and also see ``President answers to Supreme Leader''. And think to myself probably President is the highest authority in a different sense and not like being the Head of State!
    So I would trust both but try to justify them somehow! So I would come up with my own rationale: maybe Supreme Leader's authority is different. E.g. only considering religious matters.
    So when I don't have any prior knowledge I'm more likely to just trust what I see.''}[P5]
\end{myquote}

\paragraph{\textbf{Task Effects}}
The final part of our interviews contrasted different characteristics of simple (e.g., Lookup or Factoid Question Answering tasks) against complex and exploratory search tasks (e.g., Learn or Investigate categories) and how they are impacted by errors. 
In order to enable the participants to compare simple and exploratory search tasks, they were presented with a list of factoid questions on the same History and Politics topics and were asked to use the system for a few minutes to find answers to these questions. Participants were encouraged to think-aloud and share their strategy for locating the specific answers to each question while manipulating the knowledge graphs presented by the UI.

Once participants performed a few question-answering tasks they were asked to compare between the characteristics of this task and the previous essay writing tasks that they had completed and whether this interface supports these two types of tasks differently.
Alongside participant comments, data from our field notes and observations indicated that the impact of poor quality of extracted information is higher for lookup and factoid question answering tasks than in the exploratory search tasks. The sensemaking aspect of exploratory search tasks coupled with the time willing to be spent in these tasks seem to play a role in mitigating the impact of errors to some extent.

\begin{myquote}{0.2in}
    \textit{``I'm more likely to notice errors when I dig deeper, I'm investigating further and when I'm inclined to read the articles text. These are the case in the Complex Task.''} [P9]\\
    \textit{``[for a Complex task] I'm trying to get a big picture or form an idea, whereas here [in Simple task] I'm looking for specific information.
    It's like a `ray of light' and a small error like that would have completely deflected it!
    Where as for the complex task I was `shining a lot of light` so small errors could cast a little shadow but then you'll get lots of other beams of light! ''} [P8]
\end{myquote}

As a final question, participants were given a factoid question to find an answer to which was particularly designed to direct the user towards erroneous information in edge labels that contained the answer to this question. 
Observing the participants as they came across inaccurate and misleading information, whether or not they would notice it and how it would impact their judgement of utility of the information enabled us to first confirm some of the previous factors that were mentioned as leading to noticing errors (e.g., prior knowledge, expectations, etc) more reliably, and second, to better understand whether different search tasks at different levels of complexity are impacted by errors differently.

The question that was chosen for this final stage was from Politics of Iran and asked, ``Whether there is an authority in Iran which can dismiss the Supreme Leader?'' Since most of our participants, regardless of their initial knowledge of Politics of Iran had learned about the Politics of this country through their attempt at the Exploratory Politics tasks, and that the Supreme Leader of Iran seemed to be a very powerful political entity, they approached this question with higher prior knowledge as well as prior expectations and bias towards what the answer should be. As well, our participants were fully aware of presence of errors in knowledge graphs at this point in the study.

The relation label that contained the answer to this question was corrupted due to a parser error which resulted in a semi-readable sentence which would misleadingly specify ``Supreme Leader supervised Assembly of Experts.'' While the snippet corresponding to this sentence would contain the accurate information as ``Supreme Leader of Iran is elected and supervised by the Assembly of Experts.''

About half of our participants read the relation label and dismissed it as it was not providing an answer to the question. The other half, however, read the label and then clicked on $<$View More$>$ to see the snippet. Among this group some still did not notice the inconsistency between the information in the relation label and the snippet provided. 
A closer speculation of what they read and understood from these sentences led to interesting observations about how different types of cognitive bias had led to accepting erroneous information and not noticing the inconsistencies as a result of extraction errors.

In order to unpack this observation, in the next subsection, we provide some background on cognitive biases and the existing research on how these biases can impact human judgement, and in particular the performance of online information seeking activities. We end this section by reflecting on cognitive biases we observed during the conducted user study.

\paragraph{\textbf{Bias}}
Cognitive biases are defined as a pattern of deviation in judgement occurring in particular situations, where a deviation may refer to a difference from what is normatively expected, either by the judgement of people outside of the situation or by independently verifiable facts \cite{tversky74}. 
Biases play a central role in human judgement and decision making and span a number of different dimensions. As well, biases can be observed in information retrieval in situations where searchers seek or are presented with information that significantly deviates from the truth \cite{white13}. 

Prior work has identified many different types of bias. Biases are not necessarily impacting the human judgement and their decision making process negatively. For example, they can form information-processing shortcuts leading to more effective actions in a given context or enable faster decisions when timeliness is more valuable than accuracy \cite{tversky74, white13}. 
One type of bias, however, that can lead to observing irrational search behavior is the {\it confirmation bias}, which describes people's unconscious tendency to prefer confirmatory information \cite{nickerson98}. During information seeking activities, this type of bias can make searchers to seek evidence that supports their hypothesis and disregard evidence that refutes it \cite{klayman87, white13}.

As previously discussed, most participants when pressured in time and engaged in information seeking activities, tend to skim sentences and only read the keywords. These participants mentioned their eyes are only focused on the keywords while the rest of the sentence is framed in their mind. We observed that how different types of bias led this framing of the sentences to result in misjudging the utility of the information.
While `confirmation bias' led some of the participants to just accept ``Assembly of Experts'' as yet another council that is supervised by the Supreme Leader, to our surprise, even some of the participants with very high prior knowledge of this topic who already knew the answer to this question did not notice the error in the sentence. When they were asked to describe their thought process we learned that higher prior knowledge can bias the framing of the sentences in a way that should make sense to the reader:
\begin{quote}
    \textit{ I actually read the sentence as ``Supreme Leader is supervised by''! So I added `is' and `by' to the sentence myself.  Maybe because I expected this? I mean I knew that Assembly of Experts supervises Supreme Leader. So I already knew the answer and I was only looking for some evidence to support my answer. So it's like my mind is framing the sentences or reading them in a way that I expect them to be!} [P15]
\end{quote}


Research on anchoring and adjustment has shown that people typically perform little revision to their beliefs, especially if those beliefs are strongly held \cite{tversky74, white13}. 
Furthermore, it is well established that people have constraints on their ability to process information. 
In fact, information foraging theory \cite{pirolli05}, inspired by the idea of bounded rationality and satisficing, states that information seekers, will choose behaviors that optimize the utility of information gained as a function of interaction cost. 
We notice this bias more strongly in the participants who approached the exploratory Politics task with an intention of finding evidence to support the power of the president they believed is more powerful as well as in specific question answering tasks for mid-range and high prior knowledge participants.

\subsection{Discussion}
The goal of this section was to explore the impact of error prone information extraction on exploratory search tasks supported with HKGs.  From our quantitative results, we note that Group, rather than Error condition or Task, resulted in significantly different performance on exploratory search tasks, as highlighted by the dependent measure \emph{Mark}. 

To investigate this further, we looked more closely at any confounds within each group that could potentially impact the variance we see in search performance and behavior. Because of the structure of our experimental task, Group, alongside being a between subject factor (encoding differences between participant) also encodes topic to error assignment (as shown in Table \ref{effort_performance_table}).  Groups differ in which topic was assigned to Gold versus Auto in order to balance out topic effects. Given our measure of Group as a significant factor, it is possible that lower quality extractions (i.e., Auto condition) impacted the two search tasks differently. Our qualitative data provides some evidence that task was perceived differently by participants; however, we find that, while participants were split in their perception of the complexity of the tasks, one task did not dominate another in terms of complexity.

To further probe task-error effects, we performed a Univariate ANOVA analysis on History and Politics marks separately and noticed a statistically significant effect of group on History marks ($F_{3, 25} = 4.373, p < 0.05, \eta^2 = 0.4$) but not on Politics marks. This distinction indicates that the performance of our participants may have been impacted by the error condition assigned for the History task but not for the Politics task even though both  datasets were generated with comparable levels of precision and recall.
Repeating the same analysis on History marks using two fixed factors ErrorEffect and OrderEffect indicated a highly significant effect of ErrorEffect ($F_{1, 25} = 10.806, p < 0.005, \eta^2 = 0.34$). OrderEffect, however, had no statistically significant impact on History marks.

While care must be taken in over-reach with post-hoc exploratory (non-hypothesis driven) statistical analyses, our qualitative data highlights two possible explanations for this group effect. First, prior knowledge of participants could help them recover from the poor quality of the knowledge graphs in terms of the information they represent. Second, the History task was identified as a investigate-style exploratory search task, and this class of search tasks might be more impacted by recall rates than a learn-style task.

\begin{figure}[ht!]
\centering
\includegraphics[width=0.8\columnwidth]{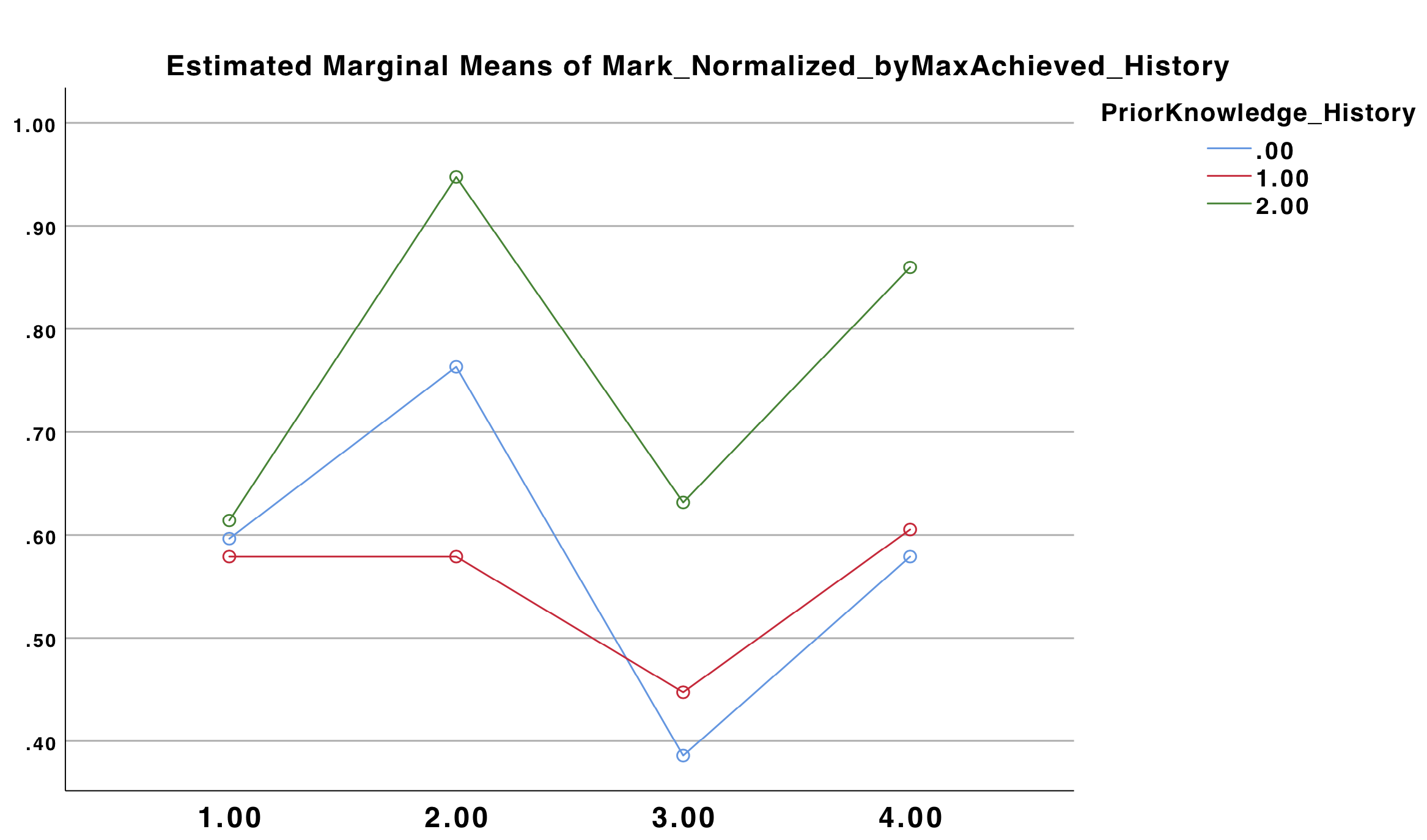}
\caption{Effects of Prior Knowledge on Marks for History Task. x-axis indicates Group as a random factor and y-axis indicates Mark as a dependent variable.}
\label{priorknowledge}
\vspace{-2mm}
\end{figure}

To explore potential knowledge versus task-style effects, we note that our pre-task questionnaires data provided an assessment of participant knowledge.  Figure \ref{priorknowledge} shows a potential correlation between prior knowledge and overall performance (i.e., Marks) on search tasks.  To test this observation, we coded prior knowledge from our three questions onto a three-point scale. 
We performed a Univariate ANOVA with ErrorEffect and OrderEffect, as fixed factors, prior knowledge as a random factor, and History marks as a dependent variable. The results indicated no statistically significant impact of prior knowledge, ErrorEffect, and OrderEffect. There were also no significant interaction between Group factors and prior knowledge.

Summarizing these observations, while we observe no initial effect of error on performance, combining qualitative data with post-hoc statistical analysis, we find some evidence that precision and recall rates may impact one of our tasks, the History task, more than the Politics task.  This, potentially makes sense; because the History task is an investigate task with, as noted qualitatively by our participants, a set of answers that are targeted rather than open-ended, errors in precision and/or recall might result in concepts useful to the search task being omitted from the data set.

\section{Synthesis, Implications, and Future Work}

Inspired by our earlier work contrasting the support networks and hierarchies provide for exploratory search, this paper explores a novel data structure, hierarchical knowledge graphs.  The goal of HKGs is to combine the complementary advantages of individual data structures.  Specifically, we observed that hierarchies provide better sensemaking for searchers new to a topic area by structuring the information space; whereas networks contain greater information within the data structure, thus reducing the need to read documents to acquire information.

HKGs are a natural extension of knowledge graphs. A knowledge graph is generated using information extraction algorithms, and then a hierarchy is generated from this knowledge graph through an iterative thresholding algorithm.  Essentially, for each entity, the `degree' of the entity (i.e., the number of edges relating it to other entities) is calculated, and thresholding selects those entities with higher degree as `central concepts.' These central concepts are then leveraged, via thresholding, to create a hierarchy within the data structure.

After presenting the design of HKGs, this paper presents two mixed methods studies of the efficacy of HKGs.  First, in a controlled experiment where underlying information extraction is accurate, we show the HKGs, quantitatively exhibit the benefits of knowledge graphs in limiting the need to read, and that the hierarchical structure of HKGs, while perceived differently than hierarchies generated via tables-of-contents, are leveraged by participants for sensemaking, and, in particular, for providing an overview of the topic domain.  Second, through a second controlled experiment, we show that HKGs exhibit resilience to errors in information extraction, an important consideration for their real-world adoption, given that natural language processing is still continually being adapted and improved for a variety of tasks (e.g., information extraction).

More broadly, this paper further argues for the complementary benefits of both hierarchical and network representations. In our earlier work, we identified benefits of hierarchies, as representations of tables-of-contents, and networks, as representations of the entity-relationship concepts within a document corpus for exploratory search, but what was unclear was whether these distinct benefits could complement each other within a single data structure encapsulating search results. Our hierarchical knowledge graphs demonstrate one feasible way that this is possible. 

Alongside this demonstration of complementary nature of networks and hierarchies, in our second experiment, we demonstrate a disassociation between \emph{output} and \emph{outcomes} within our exploratory search system. While benchmarking demonstrates error-prone information extraction, our measures of user performance demonstrate only limited impacts, and then only qualitative, of errors in information extraction (the output) with respect to user performance (the outcomes).  Obviously, there exist a number of domains where output accuracy is highly relevant: specific document lookup, legal document discovery, and research reference lookup are all examples of such.  However, it is possible that other domains, such as exploratory search, may be more resilient to errors in output accuracy, particularly if an effective interface is used to support interaction with the information such that users, while performing their search tasks, can dynamically accommodate for errors in information during sensemaking \cite{almaskari08good, sanderson10, smucker10}. 
\subsection{Limitations and Future Work}

In any study, there exist limitations.  We took significant care to address potential limitations in our study by performing power estimates in initial quantitative results to ensure a sufficient sample size for statistical effects to become visible; triangulating quantitative and qualitative data to cross-validate results; and performing post-hoc quantitative analysis to validate qualitative themes that emerged from participants. However, one challenge with measuring the effect of IE errors is that, due to the need to understand how error impacts outcomes, tasks need both an experimental condition (automatic information extraction) and a control condition (manually corrected IE). This limits the number of tasks to those with manually generated ground truth. We control for this by cross-validating our results in ground truth output with past results and leveraging the same tasks.

One challenge with extending the work to additional search queries is that, while we use automatic IE algorithms in our second study, these algorithms are still limited in performance.  While our algorithm produces automatic results, it does not do so in real-time; natural language processing for information extraction remains an algorithmically complex task.  One open question that we continue to explore is whether there exist algorithms that can run in real-time with acceptable accuracy.  However, we have yet to identify algorithms with both reasonable accuracy and sufficiently rapid performance so as to dynamically perform information extraction in response to a user's query.  This is an interesting area of future work -- to both identify real-time capable algorithms, regardless of their accuracy, and to assess the effect that these, potentially more error-prone algorithms, might have on exploratory search tasks.

Alongside the need for future work leading to real-time performance, another obvious area of future work is to add additional exploratory search tasks from Marchionini's taxonomy \cite{marchionini06}. Our emergent qualitative results in our second experiment indicate that, within the broad category of exploratory search tasks, different types of exploratory search may be impacted differently by errors. Understanding where and how the current levels of precision for IE algorithms impact each of these different task types will help to clarify where and how useful variable IE accuracy is for different types of tasks at a more fine-grained level.

This paper primarily focuses on one representation of information, HKGs. While different representations may yield different interaction behaviours, HKGs belong to the class of search systems that directly organize and present search results using spatial representations \cite{wilson10} and were designed to combine alternative information browsing paradigms \cite{shneiderman96}. 
While we cannot suggest that our results would apply to other classes of search interfaces (e.g., systems that provide a classification of the results using different metadata or facets), we do believe that our findings encourage further investigation to test both the efficacy of representations and the resistance of representation to errors in retrieval or information extraction.

Finally, beyond a focus on information representations and their generation, anther avenue of future work is to investigate other potential factors that may affect user performance in search tasks.  For example, 
cognitive biases can result in irrational search behavior and influence searchers' relevance judgment of information \cite{white13}.  As well, bounded rationality impacts the way information seekers optimize their information processing efforts even at the cost of achieving a sub-optimal outcome \cite{pirolli05}.
Our qualitative data provides some evidence that biases might be salient:  participants who approached the exploratory Politics task with an intention of finding evidence to support the power of the president they believed is more powerful limited browsing behaviors because participants felt `already informed' on the topic.  Given the difficulty in fully controlling for cognitive biases and ability, this final research question would require extensive studies to both identify factors and to model them in a tractable way to measure their effects.
\section{Conclusion}
In the field of information retrieval, alongside document retrieval, issues of representations that allow users to make sense of information and interfaces that allow users to interact with search results are important areas of inquiry. In this paper, we explore hierarchical knowledge graphs, an extension of knowledge graphs that leverages connectivity to generate hierarchies from the underlying knowledge graphs.  Our mixed method experimental results argue that hierarchical knowledge graphs support the overview advantages of hierarchical representations, the information content advantages of knowledge graphs, and exhibit resilience to information extraction error rates common in contemporary information extraction algorithms.

\bibliographystyle{ACM-Reference-Format}
\bibliography{KG4ES_Sig.bib}

\newpage
\appendix
\section{Additional Information on Experimental Design}
We provide additional details regarding the exploratory search UI as well as other resources we leveraged to be able to replicate our past works \cite{sarrafzadeh17, sarrafzadeh16}. 
\vspace{-3mm}
\subsection{Control Exploratory Search Interface}
\label{ControlUIs_details}
For the first study presented in this paper (see Section \ref{ExperimentalDesign_PartI}) we leveraged two control interfaces from our prior work \cite{sarrafzadeh16} in order to validate whether our HKGs can preserve the known strengths of network and hierarchical structures. We provide more details about these reference UIs in this section.

\begin{figure*}[htb!]
\centering
\includegraphics[width=1.09\textwidth]{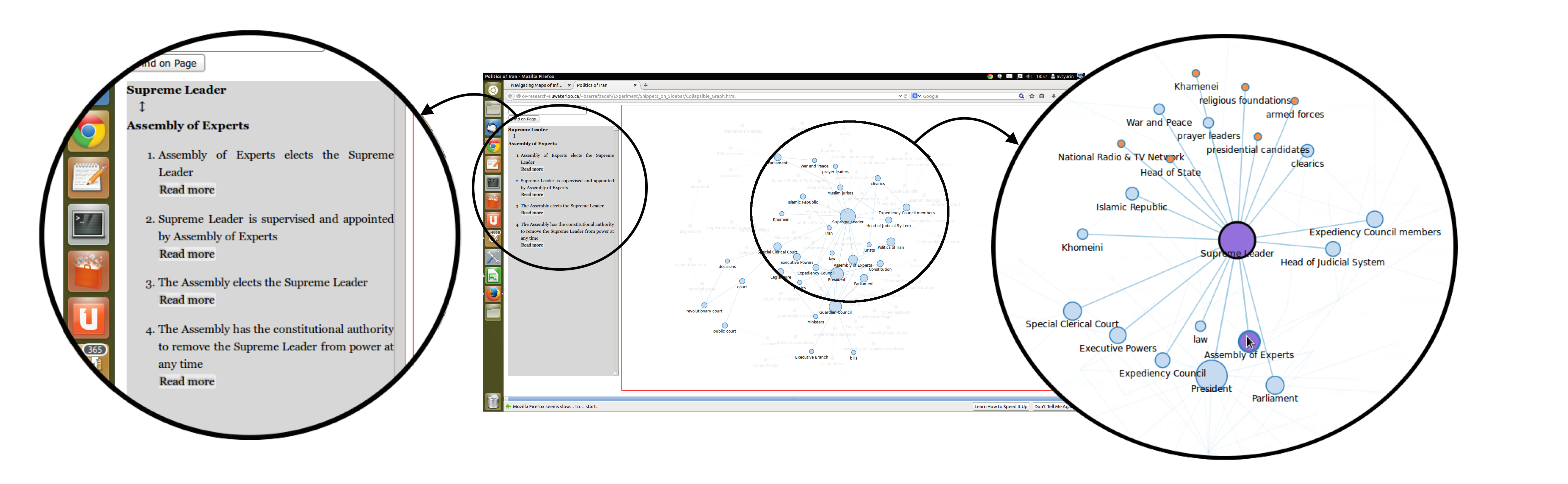}
\includegraphics[width=1.09\textwidth]{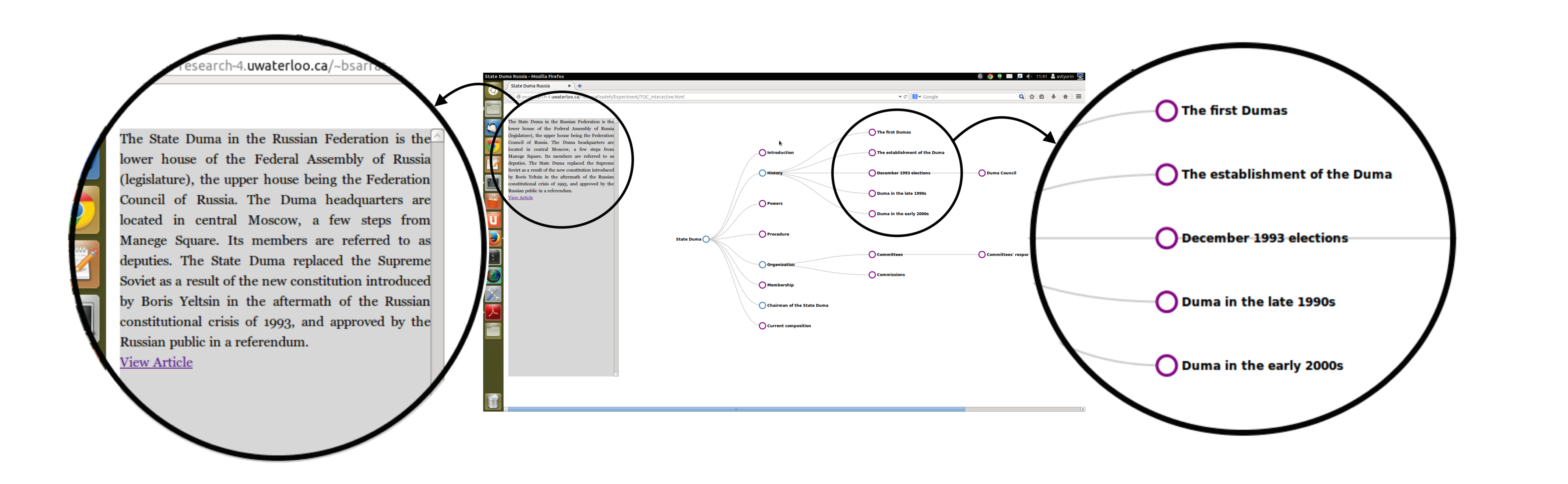}
      \caption{The graph and tree control interfaces. Note the callouts of nodes and document snippets.}
\label{Graph_UI}
\end{figure*}

The first interface employed a graph structure in which the entities from each article were the nodes and the sentences describing a semantic relation between them were the edges. 
When the user launches the graph-based application (Figure \ref{Graph_UI}, top), they are presented with a knowledge graph containing labelled nodes and unlabelled links between nodes. Nodes that represent entities with low frequency are hidden in the initial view, and only appear once a higher-frequency, connected node is clicked. This ensures that the graph does not become too cluttered. Once the user clicks on a node, that node and all connected nodes are highlighted, while the remainder of the graph is alpha-blended into the background. By hovering over any connected node in highlighted portion of the graph, the user can see the relationship(s) between the two nodes in the snippet window located on the left side of the interface (Figure \ref{Graph_UI}. top). For each relationship in the snippet region, participants have a link that allows them to view the corresponding Wikipedia article.

The second interface utilized a hierarchy (or a tree) structure to organize headings and sub-headings of the articles, as observed in each page's table-of-contents. 
The tree interface is shown in Figure \ref{Graph_UI}, bottom. When the user launches the application, the user is presented with a fully expanded tree. By clicking on any node within the tree, that portion of the wikipedia document corresponding to the node is presented in the preview area at the left of the interface. Under the snippet in question, there is a link to view article, allowing users to access the article in question.

\subsection{Study Resources}
\label{StudyResources}
\subsubsection{Questionnaires}
\label{Appendix_Questionnaires}
All demographics, pre and post task questionnaires that are used for both studies are available online through the first author's homepage\footnote{\url{https://cs.uwaterloo.ca/~bsarrafz/HKG/Experiment/Forms}}.

\subsubsection{Semi-structured Interviews.} 
\label{interviews}
In this section we provide a list of questions that guided our semi-structured interviews for both studies presented in this paper.

For the first study each participant used the HKG UI to complete one simple and one complex task (with a randomized order). Once both tasks were completed we conducted semi-structured interviews to gauge their experience with the interface, whether HKGs are more suitable for one task than the other and how they fare against our previously designed Hierarchical Tree UI (described in \ref{ControlUIs_details}). 

\begin{enumerate}
    \item Part 1: Contrasting the utility of the HKG interface for simple versus complex tasks.
    \begin{enumerate}
        \item Now that you used the same interface for two types of tasks; for which type do you find it more suitable? for question answering task or more open ended one?
        \item How about your usual Web search experience? How would you go about performing these two tasks using a search engine? How is that different from using this new interface?
        \item Would your domain knowledge change your preference about the type of task you would do with this interface?
    \end{enumerate}
    \item Part 2: Demonstrating the Hierarchical Tree based UI as a reference interface, allowing the participant to explore the features and get a \textit{feel} of how completing search tasks using the reference interface would look like.
    \begin{enumerate}
        \item Now that you did a question-answering task and an essay writing task, you can imagine you were given the other interface [Tree-based UI] instead. How would you like it? How does it compare with the [HKG] UI you used for these tasks?
        \item What are other factors that might affect your choice of interface? (e.g. the search domain, task type, prior knowledge, etc.)
        \item Are there specific scenarios where you prefer a hierarchical tree UI over the HKG interface?
    \end{enumerate}
\end{enumerate}

\vspace{5mm}

In the second study participants used the same HKG UI to complete two complex search tasks. This provided an opportunity to gauge participants' perception of the exploratory search tasks we designed and better characterize their main attributes. As well, we investigated how HKGs of varying quality were perceived by participants and how information seeking can proceed in presence of errors. What follows is the main steps that were generally followed in the interviews:
\begin{enumerate}
    \item Perception of Task Complexity: how did you feel about the complexity of these tasks? how were they different?
    \item Perceived Quality of Generated Knowledge Graphs: how did you find the quality of the information represented by the graphs? was one better than the other? did you notice any errors, inconsistencies, incomplete or unreadable sentences, missing information, etc?
    \item Notifying the participant about the experimental conditions: one of your tasks used automatically generated knowledge graphs while the other used experts-curated data. Can you guess which task was done with the automatically generated graphs?
    \item Discussing the reasons that might have hindered participants from noticing errors or lower quality of graphs.
    \item Contrasting different characteristics of simple versus exploratory search tasks and how they might be impacted by errors. This step involved using the UI for a few simple queries in order to familiarize the participant with factoid type queries.
    \item Inquiry regarding a factoid question where the relevant edge in the graphs was erroneous by design.
\end{enumerate}

\subsubsection{Marking Scheme for Essays.} 
\label{MarkingSchemes}
We followed the same experimental procedure as prior work \cite{sarrafzadeh16} were participants would engage in exploratory information seeking tasks in order to write a short essay articulating their responses for the given task. We replicated the same search tasks as described in Section \ref{SearchTasks} and leveraged the marking scheme from our past work to evaluate the quality of the essays provided by our participants.

The marking scheme for the Politics task requires 3 main arguments to be included to support one of the presidents to be more powerful than the other. In order to ensure these arguments are based on the information that was retrieved using the system (and not based on the participant's prior knowledge) each argument needs to have references to the source document. Additionally, each argument is graded as follows:
\begin{itemize}
   \item 1 point: an argument only refers to one president;
    \item 2 points: an argument compares both presidents on the same aspect (e.g. military power, rank in the political system, etc.);
    \item 3 points: same as previous, but the argument also contains information about the authority that limits the president’s power (requires a broader understanding of the political system and the power relationships between different political entities);
\end{itemize}

For the History task, the marking scheme indicates 6 cities as previous capitals of Canada and specifies the marking criteria as:
\begin{itemize}
    \item 1 point: 3 or fewer correct capitals are listed;
    \item 2 points: at least 4 correct capitals are listed;
    \item 2 points: for every well described reason behind changing a capital that is explicitly mentioned in a reference document;
    \item 1 point: for every questionable or subjective reason provided which includes a reference document;
\end{itemize}

\end{document}